\documentclass[prl,aps,amssymb,amsmath]{article}
\usepackage{amssymb,amsmath,tikz}
\usepackage{amsthm}
\usepackage{mathrsfs}
\usepackage{pifont}

\showoutput
\theoremstyle{plain}

\newtheorem*{theorem*}{THEOREM}
\newtheorem*{lemma*}{LEMMA}
\newtheorem{theorem}{THEOREM}
\newtheorem{lemma}{LEMMA}
\newtheorem{definition}{DEFINITION}
\newtheorem{definition*}{DEFINITION}
\newtheorem*{example*}{EXAMPLE}
\newtheorem*{example}{EXAMPLE}

\newtheorem*{rem*}{REMARK}

\newtheorem*{notn*}{NOTATION}
\newtheorem*{wiener-ito*}{WIENER-IT\^O-SEGAL DECOMPOSITION}

\begin{document}
\title{{\bf Bogoliubov's causal perturbative QFT with Hida operators}\footnote{Presented at the International Conference 
``Selected Topics in Mathematical Physics'',
held at Steklov Mathematical Institute, September 27-30}}
\author{Jaros{\l}aw Wawrzycki
\\
Bogoliubov Labolatory of Theoretical Physics, 
\\ Joint Institute of Nuclear Research,\\
141980 Dubna, Russia\\
\\Institute of Nuclear Physics of PAS, 
\\
ul. Radzikowskiego 152, 31-342 Krak\'ow, Poland}
\maketitle

\vspace{1cm}

\begin{abstract}
We will present the axioms of Bogoliubov's causal perturbative QFT in which the creation-
annihilation operators are interpreted as Hida operators. We will shortly present the results
that can be achieved in this theory:
1. Removal of UV and IR infinity
in the scattering operator,
2. Existence of the adiabatic limit
for interacting fields in QED,
3. Proof that charged particles
have non-zero mass,
4. Existence of infrared and ultraviolet
asymptotics for QED. 

\vspace*{1cm}

{\small {\bf keywords}: scattering operator; causal perturbative method in QFT; 
 interacting fields; white noise; Hida operators; integral kernel operators; Fock expansion.}
\end{abstract}

\section{Introduction}\label{Introduction}

It was the monograph \cite{Bogoliubov_Shirkov} where a way to the rigorous formulation of the renormalization method in perturbative QFT was initiated. The 
causal axioms (I) - (IV) (see below) were formulated in \cite{Bogoliubov_Shirkov} for the scattering operator $S$.  This enabled the perturbative QFT to be  transferred into the axiomatic path, and it was there that the idea of a strict mathematical construction of higher-order contributions $S_n$ to $S$ first appeared, 
which uses only the axioms (I) - (IV), plus, possibly, Ward's identities, cf. \cite{Bogoliubov_Shirkov}, \S 29.2. 

Some freedom, however, concerning the strict mathematical interpretation of the generalized free field operators, their Wick products
and the higher order contributions $S_n$ to $S$ was left open. Concerning this freedom, it was stated in \cite{Bogoliubov_Shirkov} only 
a general assumption that these generalized operators should be some kind of operator distributions, which can also be evaluated 
at the Grassmann valued test functions. The class of generalized operators should include the free fields
and their Wick products, and the Wick polynomials in free fields, with the coefficients equal to arbitrary translationally invariant
tempered distributions. This class should include the free fields and the Wick product operation, which is well-defined whenever 
multiplied by any translationally invariant tempered distribution. The class of generalized operators which is sufficient to include the higher 
order contributions $S_n$, is characterized in \cite{Bogoliubov_Shirkov} as consisting of Wick polynomials in free fields with the scalar coefficients
being tempered translationally invariant distributions. 
The higher order contributions $S_n$ are not uniquely determined by the causality axioms (I)-(IV), 
but only up to a quasi local generalized operators $\Lambda_n$ supported at the full diagonal. Using this freedom, 
it was explained in \cite{Bogoliubov_Shirkov}, that the scalar coefficients in $S_n$ (or Green functions), 
computed with the standard renormalization technique, determine $S_n$ which are in agreement with the causality axioms (I) - (IV).
But concerning the rigorous treatment of $S_n$ 
it was indicated in \S 29.2 only a proof of existence of $S_n$ based on the axioms (I)-(IV) 
without giving any explicit mathematically rigorous construction of $S_n$.
A mathematically rigorous and \emph{explicit construction} of $S_n$, based solely on the Bogoliubov's axioms (I) - (IV), 
and which gives the same scalar coefficients in $S_n$ as the renormalization method, 
was given later by Epstein and Glaser \cite{Epstein-Glaser}.

The content of the axioms (I) - (IV) for the $S$ operator will depend on how we mathematically interpret the free field operators, 
their Wick products, and the higher-order contributions $S_n$, as generalized operators. Thus, before giving any mathematically rigorous construction
of $S_n$, based on (I)-(IV), one has to fix the mathematical meaning of these generalized operators. Epstein and Glaser\cite{Epstein-Glaser} assumed that these operators are operator valued distributions precisely in the Wightman\cite{wig} sense. Using this interpretation 
for the generalized operators, it was shown in \cite{Epstein-Glaser}
that the standard results of the renormalization technique for the computation of $S_n$ (or rather for the scalar coefficients at the 
free-field-Wick monomials in $S_n$) 
can be rigorously reconstructed from the axioms  (I) - (IV), if we add the assumption (V) of preservation of the
singularity degree at zero of the causal coefficient distributions in the computation of their splitting into the retarded and advanced parts.
In this way, we arrive at a mathematically consistent formulation of perturbative QFT, without any UV divergences. The work \cite{Epstein-Glaser} 
is in fact a continuation of the ideas outlined in \cite{Bogoliubov_Shirkov}. 
In this approach, instead of the formal (divergent) multiplication by the step theta function
in the chronological product and then renormalization removing singular parts in the ill-defined products of distributions by the step function,  
one uses the inductive step of Epstein-Glaser in the perturbative approach, which is carried out solely within the axioms (I)-(V). 
The whole problem of construction of the higher order contributions $S_n$ is thus reduced to the problem of
splitting of causal numerical tempered distributions into their retarded and advanced parts. 
The splitting of a causal distribution is not unique and depends on the so-called singularity degree at zero (in space-time coordinates) 
of the splitted distribution -- freedom corresponding to the non-uniqueness in the ordinary renormalization procedure, and associated with the 
renormalization group freedom in the approach based on renormalization.

But some IR divergences remained for those QFT with infinite range of interaction (like QED), 
whenever one wanted to pass to the limit of the intensity of interaction function
$g$ to the constant function $1$, performed in order to get results with the physical interaction 
also in the remote part of space-time. 
This is the celebrated \emph{adiabatic limit problem}.
In some problems of QFT with infinite range of interaction (like QED), passing to this limit is unavoidable.      

We propose to improve this Bogoliubov-Epstein-Glaser perturbative QFT by reinterpreting mathematically the generalized operators 
and regard them not as the operator valued distributions
in the Wightman sense, but as the integral kernel operators in the sense of the white noise calculus \cite{obataJFA}. This allows us to solve the
adiabatic limit problem -- impossible to solve when the generalized 
operators were interpreted as operator valued distributions in the Wightman sense \cite{wig}. In this way we arrive at the theory without
any UV or IR divergences, which is mathematically consistent.  

Our presentation does not give proofs. Nonetheless, it is rigorous, written in the `definition-lemma-theorem' style. 
We emphasize the essential points, indicate the basic theorems on integral kernel operators and the extension theorems
for operators on nuclear spaces which are fundamental in the proof, and try to clarify why the integral kernel operators 
are effective in the investigation of the adiabatic limit
while the operator valued distributions in the Wightman sense are not.

\section{Free fields and their Wick products understood as integral kernel operators}
 
In other words, we are using the Hida white noise operators 
\[
\partial_{\boldsymbol{p}}^{*}, \,\,\, \partial_{\boldsymbol{p}}
\]
which respect the canonical commutation or anticommutation relations
\[
\big[\partial_{\boldsymbol{p}}, \partial_{\boldsymbol{q}}^{*}\big]_{{}_{\mp}} = \delta(\boldsymbol{p}-\boldsymbol{q}),
\]
as the creation-annihilation operators 
\[
a(\boldsymbol{p})^{+}, \,\,\, a(\boldsymbol{p}),
\]
of the free fields in the Bogoliubov's causal perturbative QFT,
leaving all the rest of the theory completely unchanged.
\emph{I.e.}
we are using the standard Gelfand triple
\[
\left. \begin{array}{ccccc} 
E & \subset & \mathcal{H} & \subset & E^* 
\end{array}\right.,
\]
over the single particle Hilbert space $\mathcal{H}$ of the total system of free fields 
determined by the corresponding standard self-adjoint operator $A$ in $\mathcal{H}$ (with some negative power $A^{-r}$ being nuclear), 
and its lifting to the standard Gelfand triple 
\[
\left. \begin{array}{ccccc} 
(E) & \subset & \Gamma(\mathcal{H}) & \subset & (E)^* 
\end{array}\right.,
\]
over the total Fock space $\Gamma(\mathcal{H})$ of the total system of free fields with the corresponding standard operator $\Gamma(A)$.
Here, $E$ is the single particle test space in the total single particle space
in the total Fock space, and is equal to the direct sum of the particular single particle test spaces $E_1, E_2, \ldots$ of the particular free fields
$\mathbb{A}^{(1)}, \mathbb{A}^{(2)}, \ldots$ undelying the QFT in question,
which together with their strong duals $E_{1}^{*}, \ldots $ and the total single particle Hilbert spaces
compose Gelfand triples. (The respective single particle test spaces, the direct summands of $E$,
of the corresponding free fields $\mathbb{A}^{(1)}, \mathbb{A}^{(2)}, \ldots$,  will be denoted by 
$E_1, E_2, \ldots$ or $E_{{}_{'}}, E_{{}_{''}}, \ldots$. The plane wave kernels of the free fields will be denoted
by $\kappa_{0,1}^{{}^{(1)}}$, $\kappa_{1,0}^{{}^{(1)}}, \ldots$ or 
$\kappa'_{0,1}$, $\kappa'_{1,0}$, $\kappa''_{0,1}$, $\kappa''_{1,0}\ldots$.) 
\[
\left. \begin{array}{ccccc} 
E & \subset & \mathcal{H} & \subset & E^* \\
\parallel & & \parallel & & \parallel \\
E_1 \oplus \ldots \oplus E_N & & \mathcal{H}_1 \oplus \ldots \oplus \mathcal{H}_N & & E_{1}^{*} \oplus \ldots \oplus E_{N}^{*}
\end{array}\right.,
\]
In general the nuclear spaces $E_1, E_2, \ldots$ are naturally defined as linear spaces of smooth sections of smooth vector bundles, but 
are unitary isomorphic (through a natural unitary isomorphism, defined through the smooth idempotent associated to the smooth bundle 
$E_1, E_2, \ldots$, and which is also continuous in the nuclear topologies) to the
nuclear spaces of test $\mathbb{C}^{d_1}$, $\mathbb{C}^{d_2}$, $\ldots$-valued functions on the corresponding orbits 
$\mathcal{O} = \{p: p\cdot p =m, p_0\geq 0  \}$ in momentum space for the fields of mass $m$, which can be represented as test function
spaces  $E_1 = \mathcal{S}(\mathbb{R}^3; \mathbb{C}^{d_1}), \ldots$ in massive case, or as  
$E_1 = \mathcal{S}^{0}(\mathbb{R}^3; \mathbb{C}^{d_1}), \ldots$, in massless case $m=0$, and are restrictions to the correponding orbits
$\mathcal{O}$ of the Fourier transforms of the scalar, vector, spinor, $\ldots$, (depending on the kind of field $\mathbb{A}^{(1)}, \ldots$) 
functions lying, respectively, in the Fourier inverse images of   $\mathcal{S}(\mathbb{R}^4; \mathbb{C}^{d_1}), \ldots$ or
$\mathcal{S}^{0}(\mathbb{R}^4; \mathbb{C}^{d_1}), \ldots$. Here $\mathcal{S}^{0}(\mathbb{R}^n; \mathbb{C}^{d})$ is the closed subspace
of $\mathcal{S}(\mathbb{R}^n; \mathbb{C}^{d})$ of all those functions who's all derivatives vanish at zero:
\[
\left. \begin{array}{ccccc} \mathcal{S}_{A}(\sqcup \mathbb{R}^3; \mathbb{C}) & \subset & L^2(\sqcup \mathbb{R}^3;\mathbb{C}) & \subset & \mathcal{S}_{A}(\sqcup \mathbb{R}^3; \mathbb{C})^* \\
\downarrow \uparrow & & \downarrow \uparrow & & \downarrow \uparrow \\
E & \subset & \mathcal{H} & \subset & E^*
\end{array}\right..
\]

This means that the space $E$ consisting of direct sums of restrictions of the Fourier transforms of space-time test $\mathbb{C}^d$-valued 
functions (or rather smooth sections -- their images under a smooth idempotent in case of higher spin charged fields)
to the respective orbits $\mathcal{O}$ in momenta defining the representation of $T_4 \ltimes SL(2, \mathbb{C})$ in the Mackey's classification
(acting in the single particle Hilbert space for each corresponding free field) is given the standard realization with the help of a standard operator
$A$ in $L^2(\sqcup \mathbb{R}^3; \mathbb{C}) \cong \mathcal{H}$ (with standard $A$, \emph{i.e.} 
self-adjoint positive, with some negative power of which being nuclear, or trace-class,
and with the minimal spectral value greater than $1$). It is equal to the 
the direct sum $\oplus_{i=1}^{N} A_i$ of the standard operators $A_i$ corresponding to the single particle Hilbert space
of the $i$-th free field. Thus, first we need to construct the standard $A_i$ and the standard Gelfand triples for each of the free fields
of the theory.  $A= \oplus_{i=1}^{N} A_i$ is also standard.  It serves for the construction of the standard Gelfand
triple over the full single particle Hilbert space $L^2(\sqcup \mathbb{R}^3; \mathbb{C}) \cong \mathcal{H}$. 
Namely, with the help of the operator $A$ we construct $E$ as the following projective limit
\[
E = \underset{k\in \mathbb{N}}{\bigcap} \textrm{Dom} \, A^k
\]
with the standard realization of $E$ as the countably Hilbert nuclear (in the sense of Grothendieck) test space and its strong (also
nuclear) dual as the inductive limit
\[
E^* = \underset{k\in \mathbb{N}}{\bigcup} \textrm{Dom} A^{-k}
\]
with the Hilbertian defining norms
\[
\big| \cdot \big|_{{}_{k}} \overset{\textrm{df}}{=} \big|A^k \cdot \big|_{{}_{L^2}}, \,\,\,\,\,\,
\big| \cdot \big|_{{}_{-k}} \overset{\textrm{df}}{=} \big|A^{-k} \cdot \big|_{{}_{L^2}}, \,\,\, k = 0,1,2,3, \ldots.
\]

For any $\Phi$ in $(E)$ or in $(E)^*$ let
\[
\Phi = \sum\limits_{n=0}^{\infty} \Phi_{n} \,\,\,\,
\textrm{with} \,\,\, \Phi_n \in E^{\hat{\otimes} \, n} \,\, \textrm{or, respectively}, \,\,\, \Phi_n \in E^{*\hat{\otimes} \, n}
\]
be its decomposition into $n$-particle states of an element $\Phi$ of the test Hida space $(E)$
or in its strong dual $(E)^*$, convergent, respectively, in $(E)$ or in $(E)^*$. We define
\[
\begin{split}
a(w) \Phi_0 = 0, \,\,\,\, a(w) \Phi_n = n \, \overline{w} \hat{\otimes}_1 \Phi_n
\\
a(w)^+ \Phi_n = w \hat{\otimes} \Phi_n, \,\,\,\, \textrm{for each fixed} \,\, w \in E^*.
\end{split}
\]
\begin{definition}
The Hida operators are obtained when we put here the Dirac delta functional $\delta_{{}_{s,\boldsymbol{p}}}$ for $w$
\[
\partial_{{}_{s,\boldsymbol{p}}} = a_s(\boldsymbol{p}) = a(\delta_{{}_{s,\boldsymbol{p}}}),
\,\,\,\,
\partial_{{}_{s,\boldsymbol{p}}}^+ = a_s(\boldsymbol{p})^+ =  a(\delta_{{}_{s,\boldsymbol{p}}})^+.
\]
\label{Def:HidaOperators}
\end{definition}

Let $L(E_1,E_2)$ be the linear space of linear continuous operators $E_1\longrightarrow E_2$ endowed with the natural topology of uniform
convergence on bounded sets. 

For each fixed spin-momentum point $(s,\boldsymbol{p})$ the Hida operators are well-defined
(generalized) operators 
\[
\begin{split}
a_s(\boldsymbol{p}) \in L \big((E), (E)\big) \subset L \big((E), (E)^*\big),
\\
a_s(\boldsymbol{p})^+ \in L \big((E)^*, (E)^*\big) \subset L \big((E), (E)^*\big),
\end{split}
\]
with the last ``$\subset$'' by topological inclusion $(E) \subset (E)^*$. 
Let $\phi \in \mathcal{E}$ (here $\mathcal{E}$ is the space-time test space $\mathcal{S}$ or $\mathcal{S}^{00}$) 
and let $\kappa_{l, m}$ be any $L(\mathcal{E}, \mathbb{C}) = \mathcal{E}^*$-valued distribution
\[
\kappa_{l, m} \in  L (E^{\hat{\otimes} (l+m)}, \mathcal{E}^*)  
= L(\mathcal{E}, E^{*\hat{\otimes} (l+m)}) = E^{* \hat{\otimes} (l+m)} \otimes \mathcal{E}^*, 
\]
then we put
\begin{definition}
\[
\left. \begin{array}{ccc} 
\Xi_{{}_{l,m}}(\kappa_{{}_{l,m}})(\Phi \otimes \phi) & \overset{\textrm{df}}{=}  &
\sum\limits_{n=0}^{\infty} \kappa_{l,m} \otimes_m (\Phi_{n+m} \otimes \phi)  \\
\parallel & & \parallel   \\
\Xi_{{}_{l,m}}\big(\kappa_{{}_{l,m}}(\phi)\big) \, \Phi &  &  \sum\limits_{n=0}^{\infty} \kappa_{l,m}(\phi) \otimes_m \Phi_{n+m}
\end{array}\right.,
\]
\label{Def:Xi}
\end{definition}
which for any $\mathcal{E}^*$-valued distribution $\kappa_{l,m}$ is a well-defined (generalized) operator 
\begin{multline*}
\Xi(\kappa_{lm}) = 
\int \kappa_{lm}\big(\boldsymbol{p}_1, \ldots, \boldsymbol{p}_l,
\boldsymbol{q}_1, \ldots, \boldsymbol{q}_m\big) \,\,
\partial_{\boldsymbol{p}_1}^{*} \ldots \partial_{\boldsymbol{p}_l}^{*} 
\partial_{\boldsymbol{q}_1} \ldots \partial_{\boldsymbol{q}_m} \, \times
\\
\times \,
d\boldsymbol{p}_1 \ldots d\boldsymbol{p}_l
d\boldsymbol{q}_1 \ldots d\boldsymbol{q}_m,
\end{multline*}
\[
\Xi_{{}_{l,m}}(\kappa_{{}_{l,m}}) \in L\big((E)\otimes \mathcal{E}, (E)^* \big) \cong
L\big(\mathcal{E}, L((E),(E)^*)\big).
\]
(for brevity of notation $\boldsymbol{p}_i$ denote in this formula 
spin-momentum variables  $(s_i,\boldsymbol{p}_i)$ and integrations include summations with respect to spin components
$s_i$.)

$\Xi_{{}_{l,m}}(\kappa_{{}_{l,m}})$ defines integral kernel operator $\Xi_{{}_{l,m}}(\kappa_{{}_{l,m}})$ which is uniquely determined by the condition
\[
\langle\langle \Xi_{{}_{l,m}}(\kappa_{{}_{l,m}})(\Phi \otimes \phi), \Psi \rangle\rangle
= \langle \kappa_{{}_{l,m}}(\eta_{{}_{\Phi,\Psi}}), \phi \rangle,
\,\,\,
\Phi, \Psi \in (E), \phi \in \mathcal{E}
\] 
or, respectively,
\[
\langle\langle \Xi_{{}_{l,m}}(\kappa_{{}_{l,m}})(\Phi \otimes \phi), \Psi \rangle\rangle
= \langle \kappa_{{}_{l,m}}(\phi), \eta_{{}_{\Phi,\Psi}}\rangle,
\,\,\,
\Phi, \Psi \in (E), \phi \in \mathcal{E},
\] 
depending on $\kappa_{{}_{l,m}}$ is regarded as an element of
\[
L(E^{\hat{\otimes} (l+m)}, \mathcal{E}^*)  
\,\,\,
\textrm{or, respectively, of} \,\,\,
L(\mathcal{E}, E^{*\hat{\otimes} (l+m)}).
\]
Here 
\begin{multline*}
\eta_{{}_{\Phi,\Psi}}(s_1, \boldsymbol{p}_1, \ldots, s_l, \boldsymbol{p}_l,
s_{l+1}, \boldsymbol{p}_{l+1}, \ldots, s_{l+m}, \boldsymbol{p}_{l+m})
\\
\overset{\textrm{df}}{=}
\langle\langle a_{s_1}(\boldsymbol{p}_{1})^+ \ldots a_{s_l}(\boldsymbol{p}_{l})^+ a_{s_{l+1}}(\boldsymbol{p}_{l+1})
\ldots a_{s_{l+m}}(\boldsymbol{p}_{l+m})\Phi, \Psi \rangle\rangle
\end{multline*}
is the function which always belongs to $E^{\hat{\otimes}(l+m)}$. (Hida, Obata, Sait\^o) 

\begin{example}
Free fields $\mathbb{A}$ are sums of two integral kernel operators
\[
\mathbb{A}(\phi) = \mathbb{A}^{(-)}(\phi) + \mathbb{A}^{(+)}(\phi) = \Xi(\kappa_{0,1}(\phi)) +
\Xi(\kappa_{1,0}(\phi))
\]
with the integral kernels $\kappa_{l,m}$ represented by ordinary functions:
\[
\begin{split}
\kappa_{0,1}(\nu, \boldsymbol{p}; \mu, x) =
{\textstyle\frac{g_{\nu \mu}}{(2\pi)^{3/2}\sqrt{2p^0(\boldsymbol{p})}}}
e^{-ip\cdot x}, \,\,\,\,\,\,
p = (|p_0(\boldsymbol{p})|, \boldsymbol{p}), \, p\cdot p=0, \\
\kappa_{1,0}(\nu, \boldsymbol{p}; \mu, x) =
{\textstyle\frac{g_{\nu \mu}}{(2\pi)^{3/2}\sqrt{2p^0(\boldsymbol{p})}}}
e^{ip\cdot x},
\,\,\,\,\,\,
p \cdot p = 0,
\end{split}
\]
for the free e.m.potential field $\mathbb{A}=A$ (in the Gupta-Bleuler gauge) and
\begin{multline*}
\kappa_{0,1}(s, \boldsymbol{p}; a,x) 
\\
= \left\{ \begin{array}{ll}
(2\pi)^{-3/2}u_{s}^{a}(\boldsymbol{p})e^{-ip\cdot x}, \,\,\, \textrm{$p = (|p_0(\boldsymbol{p})|, \boldsymbol{p}), \, p \cdot p = m^2$} & \textrm{if $s=1,2$}
\\
0 & \textrm{if $s=3,4$}
\end{array} \right.,
\end{multline*}
\[
\kappa_{1,0}(s, \boldsymbol{p}; a,x) = \left\{ \begin{array}{ll}
0 & \textrm{if $s=1,2$}
\\
(2\pi)^{-3/2} v_{s-2}^{a}(\boldsymbol{p})e^{ip\cdot x}, \,\,\, \textrm{$p \cdot p = m^2$} & \textrm{if $s=3,4$}
\end{array} \right.
\]
for the free Dirac spinor field $\mathbb{A}=\boldsymbol{\psi}$, and
which are in fact the respective plane wave solutions of d'Alembert and of Dirac equation, which span the corresponding generalized eigen-solution sub spaces. 
Here $g_{\nu\mu}$ are the components of the space-time Minkowski metric tensor, and 
\[
%\begin{multline*}
u_s(\boldsymbol{p}) =  \frac{1}{\sqrt{2}} \sqrt{\frac{E(\boldsymbol{p}) + m}{2 E(\boldsymbol{p})}}
\left( \begin{array}{c}   \chi_s + \frac{\boldsymbol{p} \cdot \boldsymbol{\sigma}}{E(\boldsymbol{p}) + m} \chi_s
\\                                           
              \chi_s - \frac{\boldsymbol{p} \cdot \boldsymbol{\sigma}}{E(\boldsymbol{p}) + m} \chi_s                         \end{array}\right),
\,\,\,\,  
%=
%\frac{1}{\sqrt{2}} \sqrt{\frac{E(\boldsymbol{\p}) + m}{2 E(\boldsymbol{\p})}}
%\left( \begin{array}{c}   \chi_s + \frac{\boldsymbol{\p} \cdot \boldsymbol{\sigma}}{E(\boldsymbol{\p}) + m} \chi_s
%\\                                           
%              \beta\big(p_0(\boldsymbol{\p}), \boldsymbol{\p}\big)^2 \, \big(\chi_s + \frac{\boldsymbol{\p} \cdot \boldsymbol{\sigma}}{E(\boldsymbol{\p}) + m} \chi_s\big)                         \end{array}\right),
%\\
\]
\[
v_s(\boldsymbol{p}) =  \frac{1}{\sqrt{2}} \sqrt{\frac{E(\boldsymbol{p}) + m}{2 E(\boldsymbol{p})}}
\left( \begin{array}{c}   \chi_s + \frac{\boldsymbol{p} \cdot \boldsymbol{\sigma}}{E(\boldsymbol{p}) + m} \chi_s
\\                                           
              -\big(\chi_s - \frac{\boldsymbol{p} \cdot \boldsymbol{\sigma}}{E(\boldsymbol{p}) + m}\chi_s \big)                          \end{array}\right),
%= 
%\frac{1}{\sqrt{2}} \sqrt{\frac{E(\boldsymbol{\p}) + m}{2 E(\boldsymbol{\p})}}
%\left( \begin{array}{c}   \chi_s + \frac{\boldsymbol{\p} \cdot \boldsymbol{\sigma}}{E(\boldsymbol{\p}) + m} \chi_s
%\\                                           
%              -\beta(p_0\big(\boldsymbol{\p}), -\boldsymbol{\p}\big)^{2} \, \big(\chi_s + \frac{\boldsymbol{\p} \cdot \boldsymbol{\sigma}}{E(\boldsymbol{\p}) + m}\chi_s \big)                          \end{array}\right)
%\end{multline*}
\]
where
\[
\chi_1 = \left( \begin{array}{c} 1  \\
                                                  0 \end{array}\right), \,\,\,\,\,
\chi_2 = \left( \begin{array}{c} 0  \\
                                                  1 \end{array}\right), E(\boldsymbol{p}) = |p_0(\boldsymbol{p})|,
\]
are the Fourier transforms of the complete system of the free Dirac equation in the chiral represenation in which 
\[
\gamma^0 = \left( \begin{array}{cc}   0 &  \mathbf{1}_2  \\
                                           
                                                   \mathbf{1}_2              & 0 \end{array}\right), \,\,\,\,
\gamma^k = \left( \begin{array}{cc}   0 &  -\sigma_k  \\
                                           
                                                   \sigma_k             & 0 \end{array}\right),
\]
with the Pauli matrices
\[
\boldsymbol{\sigma} = (\sigma_1,\sigma_2,\sigma_3) = 
\left( \,\, \left( \begin{array}{cc} 0 & 1 \\

1 & 0 \end{array}\right), 
\,\,\,\,\,
\left( \begin{array}{cc} 0 & -i \\

i & 0 \end{array}\right),
\,\,\,
\left( \begin{array}{cc} 1 & 0 \\

0 & -1 \end{array}\right)
\,\,
\right).
\]
\qed
\label{KernelsForAandPsi}
\end{example}

\begin{lemma}
The free e.m. potential and spinor field operators $\mathbb{A}=A,\boldsymbol{\psi}$ are operator-valued distributions, \emph{i.e.} belong 
to $L\big(\mathcal{E}, L((E),(E))\big)$, \emph{i.e.}
\[
\kappa_{0,1}, \kappa_{1,0} \in L(E^*, \mathcal{E}^{*}) = L(\mathcal{E}, E) \subset L(E, \mathcal{E}^{*})
\cong E^{*} \otimes \mathcal{E}^{*},
\]
if and only if
\[
\begin{split}
E= \mathcal{S}^{0}(\mathbb{R}^3; \mathbb{C}^{4}) = \mathcal{S}_{A_{(3)}}(\mathbb{R}^3;
\mathbb{C}^4), \,\, \mathcal{E} = \mathcal{S}^{00}(\mathbb{R}^4; \mathbb{C}^4) = \mathcal{S}_{\widetilde{A}_{(4)}}(\mathbb{R}^4;
\mathbb{C}^4)
\,\,\,\, \textrm{for} \,\,\, A
\\
E= \mathcal{S}(\mathbb{R}^3; \mathbb{C}^4) = \mathcal{S}_{\widetilde{H}_{(3)}}(\mathbb{R}^3;
\mathbb{C}^4), \,\, \mathcal{E} = \mathcal{S}(\mathbb{R}^4; \mathbb{C}^4) = \mathcal{S}_{H_{(4)}}(\mathbb{R}^4; \mathbb{C}^4)
\,\,\,\, \textrm{for} \,\,\, \boldsymbol{\psi}.
\end{split}
\]
In this case, moreover,
\[
\kappa_{0,1}, \kappa_{1,0} \in L(E, \mathcal{O}_M)
\]
and for each $\xi \in E$, $\kappa_{0,1}(\xi), \kappa_{1,0}(\xi)$ are smooth having all derivatives bounded.
\qed
\label{Cont.free.field.kernels}
\end{lemma}

Recall that here $\mathcal{S}^{0}(\mathbb{R}^n;\mathbb{C})=\mathcal{S}_{A_{(n)}}(\mathbb{R}^n;\mathbb{C})$
is the closed subspace of the Schwartz space $\mathcal{S}(\mathbb{R}^n;\mathbb{C})$ of all functions
whose all derivatives vanish at zero. $\mathcal{S}^{00}(\mathbb{R}^n;\mathbb{C})$ is the Fourier transform
inverse image of $\mathcal{S}^{0}(\mathbb{R}^n;\mathbb{C})$. The space $\mathcal{S}^{0}(\mathbb{R}^n;\mathbb{C})$ can be  
realized as a countably Hilbert nuclear space $\mathcal{S}_{A_{(n)}}(\mathbb{R}^n;\mathbb{C})$ associated, in the sense of \cite{GelfandIV},
\cite{obata}, with a positive self adjoint operator $A_{(n)}$ in $L^2(\mathbb{R}^n; d^np)$, with $\textrm{Inf Spec} \, A_{(n)} >1$
whose some negative power $\big[A_{(n)} \big]^{-r}$, $r>0$, is of Hilbert-Schmidt class. In order to  construct an example of a series
of operators $A_{(n)}$, $n=2,3, \ldots$, let us consider the Hamiltonian operators 
\[
H_{(n)} = - \Delta_{{}_{\mathbb{R}^n}} + r^2 +1,
\,\,\,\,\,\,\,\, r^2= (p_{1})^2 + \ldots +(p_{n})^2,
\,\,\,\,\,\,\,
\textrm{in}
\,\,
L^2(\mathbb{R}^n, d^np),
\] 
and the following $L^2(\mathbb{R}\times \mathbb{S}^{n-1}; dt\times d\mu_{{}_{\mathbb{S}^{n-1}}})$,
$L^2(\mathbb{R}\times \mathbb{S}^{n-1}; \nu_{{}_{n}}(t) dt\times d\mu_{{}_{\mathbb{S}^{n-1}}})$ -spaces
on $\mathbb{R}\times \mathbb{S}^{n-1}$ with the weights
\[
\nu_{{}_{n}}(t) = {\textstyle\frac{(t+\sqrt{t^2+4})^{n-1}}{2^{n-2}(t^2+4-t\sqrt{t^2+4})}},
\] 
and the following unitary operators $U_2: L^2(\mathbb{R}\times \mathbb{S}^{n-1}; \nu_{{}_{n}}(t) dt\times d\mu_{{}_{\mathbb{S}^{n-1}}})
\rightarrow L^2(\mathbb{R}^n; d^n p)$, $U_1: L^2(\mathbb{R}\times \mathbb{S}^{n-1}; dt\times d\mu_{{}_{\mathbb{S}^{n-1}}})
\rightarrow L^2(\mathbb{R}\times \mathbb{S}^{n-1}; \nu_{{}_{n}}(t) dt\times d\mu_{{}_{\mathbb{S}^{n-1}}})$,
$U= U_2U_1: L^2(\mathbb{R}\times \mathbb{S}^{n-1}; dt\times d\mu_{{}_{\mathbb{S}^{n-1}}}) \rightarrow L^2(\mathbb{R}^n; d^n p)$
given by the following formulas
\begin{gather*}
U_1 f(t,\omega) = {\textstyle\frac{1}{\sqrt{\nu_{{}_{n}}(t)}}}f(t,\omega), 
\,\,\,\,\,\,\,\,\,\,\, f \in L^2(\mathbb{R}\times \mathbb{S}^{n-1}; dt\times d\mu_{{}_{\mathbb{S}^{n-1}}}),
\\
U_2 f(r,\omega) = f(t(r),\omega), 
\,\,\,\,\,\,\,\,\,
t(r) = r-r^{-1},
\,\,\,\,\,\,\,\,\,
f\in L^2(\mathbb{R}\times \mathbb{S}^{n-1}; \nu_{{}_{n}}(t) dt \times d\mu_{{}_{\mathbb{S}^{n-1}}}),
\end{gather*}
where $U_2 f$ is expressed in spherical coordinates $(r,\omega)$ in $\mathbb{R}^n$. We can put
\[
A_{(n)} = U \big(H_{(1)}\otimes \boldsymbol{1} + \boldsymbol{1} \otimes \Delta_{{}_{\mathbb{S}^{n-1}}} \big) U^{-1}.
\]
Here $\Delta_{{}_{\mathbb{R}^n}}, \Delta_{{}_{\mathbb{S}^{n-1}}}$ are the standard Laplace operators on $\mathbb{R}^n$
and $\mathbb{S}^{n-1}$, with the standard invariant measures $d^np, d\mu_{{}_{\mathbb{S}^{n-1}}}$. In particular
\[
A_{(3)} = - {\textstyle\frac{r^2}{r^2+1}} \partial_{r}^{2} - {\textstyle\frac{r^3(r^2+4)}{(r^2+1})^3} \partial_{r}
+ \Big[{\textstyle\frac{r^2(r^2+4)(r^2-2)}{4(r^2+1)^4}} + r^2+r^{-2} \Big]  
\]
in spherical coordinates. 

Let $\kappa^{1}_{0,1}, \kappa_{1,0}$, $\kappa^{2}_{0,1}, \kappa^{2}_{1,0}$, $\ldots, \kappa^{n}_{0,1}, \kappa^{n}_{1,0}$
be the plain wave kernels of the free fields $\mathbb{A}^{(1)}, \ldots, \mathbb{A}^{(n)}$. Then 
\[
\kappa_{lm} = \kappa^{1}_{l_1m_1} \dot{\otimes} \ldots \dot{\otimes} \kappa^{n}_{l_n m_n},
\,\,\,\, l = l_1+ \ldots + l_n, \,\, m = m_1+ \ldots + m_n 
\]
are the kernels of the Wick (pointwise) product operator
\begin{equation}\label{PointWiseWick}
{:}\mathbb{A}^{(1)}(x) \ldots \mathbb{A}^{(n)}(x) {:},
\end{equation} 
where $\dot{\otimes}$ denotes ordinary pointwise product with respect to the space time point $x$. The kernels
$\kappa_{lm}$ so defined  should be symmetrized in Boson spin-momentum
variables and antisymmetrized in the Fermion spin-momentum variables
in order to keep one-to-one correspondence between the kernels and operators. For the (tensor) Wick product operator
\[
{:}\mathbb{A}^{(1)}(x_1) \ldots \mathbb{A}^{(n)}(x_n) {:},
\] 
with $n$ independent space-time variables  $x_j$, we have analogous formula for its kernels $\kappa_{lm}$, but with the
pointwise product $\dot{\otimes}$ replaced with ordinary tensor product $\otimes$. In this case the (tensor)
Wick product operator belongs to the class $L\big(\mathcal{E}^{\otimes \, n}, \, L((E),(E))\big)$, irrespectively if the 
free fields $\mathbb{A}^{(j)}$ are massive or massless. 

We have the following easily verified lemma for the Wick (pointwise) product operator
\begin{lemma}
\[
{:}\mathbb{A}^{(1)} \ldots \mathbb{A}^{(n)} {:} \in
\begin{cases}
L\big(\mathcal{E}, \, L((E),(E))\big), 
& \text{if all fields $\mathbb{A}^{(j)}$ are massive},\\
L\big(\mathcal{E}, \, L((E),(E)^*)\big), & \text{if some $\mathbb{A}^{(j)}$ are massless fields}.\\
\end{cases}
\]
\qed
\label{WickProduct}
\end{lemma}

\begin{theorem}
The standard Wick theorem decomposition holds for the (tensor) product operator
\[
{:}\mathbb{A}^{(1)}(x) \ldots \mathbb{A}^{(n)}(x) {:} {:}\mathbb{A}^{(n+1)}(y) \ldots \mathbb{A}^{(n+k)}(y) {:}
\]
with the kernels of the decomposition given by the contractions
\[
\kappa_{{}_{l,m}}(\phi\otimes \varphi) = \sum\limits_{\kappa'_{{}_{l',m'}},\kappa''_{{}_{l'',m''}}, k}
\kappa'_{{}_{l',m'}}(\phi) \otimes_{k} \kappa''_{{}_{l'',m''}}(\varphi)
\]
where in this sum $\kappa'_{{}_{l',m'}}$ and $\kappa''_{{}_{l'',m''}}$ range over the kernels respectively
of the operators
\[
{:}\mathbb{A}^{(1)}(x) \ldots \mathbb{A}^{(n)}(x) {:} \,\,\, \textrm{and} \,\,\, {:}\mathbb{A}^{(n+1)}(y) \ldots \mathbb{A}^{(n+k)}(y) {:}
\]
and
\[
l'+l''-k=l, \,\,\, m'+m''-k=m
\]
and where the contractios $\otimes_k$ are performed upon all $k$ pairs of spin-momenta variables in which the first variable in the pair
corresponds to an annihilation operator variable and the second one to the creation operator variable
or \emph{vice versa}. All these contractions are given by absolutely convergent sums/integrals with respect 
to the contracted variables.  After the contraction, 
the kernels should be symmetrized in Boson spin-momentum
variables and antisymmetrized in the Fermion spin-momentum variables
in order to keep one-to-one correspondence between the kernels and operators.
\qed
\label{WickDecomposition}
\end{theorem}
Let us give few words of explanation. The pointwise Wick poroduct factors ${:}\mathbb{A}^{(1)} \ldots \mathbb{A}^{(n)} {:}$
and ${:}\mathbb{A}^{(n+1)} \ldots \mathbb{A}^{(n+k)} {:}$ of the last theorem, when evaluated at the test function $\phi \in \mathcal{E}$
and $\varphi \in \mathcal{E}$, are ordinary operators transforming the Hida space $(E)$ into itself, and as such
can be composed. In this case the (tesnor) product operator is well-defined, and its evaluation at the test function 
$\phi \otimes \varphi$ is, by definition, understood as equal to the composition 
\begin{equation}\label{composition}
{:}\mathbb{A}^{(1)} \ldots \mathbb{A}^{(n)} {:}(\phi) \circ {:}\mathbb{A}^{(n+1)} \ldots \mathbb{A}^{(n+k)} {:}(\varphi).
\end{equation}
If among $\mathbb{A}^{(j)}$ in ${:}\mathbb{A}^{(n+1)} \ldots \mathbb{A}^{(n+k)} {:}$ there are massless fields, then, according to
lemma \ref{WickProduct}, ${:}\mathbb{A}^{(n+1)} \ldots \mathbb{A}^{(n+k)} {:}(\varphi)$ is a generalized operator transforming
the Hida space $(E)$ into its dual $(E)^*$, and the composition (\ref{composition}) is in general meaningless.  
In this case the four-momentum $p$ in the exponent $e^{\mp ip\cdot x}$ defining the plane wave kernels
$\kappa_{0,1}, \kappa_{1,0}$ of the massless free field $\mathbb{A}^{(j)}$, ranges over the corresponding massless orbit 
$\mathcal{O} = \{p: p \cdot p =0 \}$ with zero component $p_0$ of $p$ being the following
function $p_0(\boldsymbol{p}) = | \boldsymbol{p}|$ of the spatial momentum $\boldsymbol{p}$. We replace all the massless
exponents $e^{\mp ip\cdot x}$ in $\kappa_{0,1}, \kappa_{1,0}$ of the massless fields $\mathbb{A}^{(j)}$ 
by the massive exponents with mass $\epsilon$ and $p_0(\boldsymbol{p}) = \sqrt{| \boldsymbol{p}|^2 + \epsilon^2}$. 
After this replacement both operators
in (\ref{composition}) become ordinary operators in the Fock space transformng continuously $(E)$ into $(E)$
and the composition (\ref{composition}) is meaningfull. Then the limit $\epsilon \rightarrow 0$ 
in (\ref{composition}) defines the evaluation of the (tensor) product operator at the test function $\phi \otimes \varphi$
in case where some $\mathbb{A}^{(j)}$ are massless. In general 
\begin{multline*}
{:}\mathbb{A}^{(1)} \ldots \mathbb{A}^{(n)} {:} {:}\mathbb{A}^{(n+1)} \ldots \mathbb{A}^{(n+k)} {:} 
\\
\in
\begin{cases}
L\big(\mathcal{E}^{\otimes \, 2}, \, L((E),(E))\big), 
& \text{if all fields $\mathbb{A}^{(j)}$ are massive},\\
L\big(\mathcal{E}^{\otimes \, 2}, \, L((E),(E)^*)\big), & \text{if some $\mathbb{A}^{(j)}$ are massless fields}.\\
\end{cases}
\end{multline*}
where by
\[
{:}\mathbb{A}^{(1)} \ldots \mathbb{A}^{(n)} {:} {:}\mathbb{A}^{(n+1)} \ldots \mathbb{A}^{(n+k)} {:} 
\]
we have denoted the (tensor) product operator of the Wick (pointwise) products.  

Accordingly, in the proof of theorem \ref{WickDecomposition} we proceed in two steps. In the first step, we assume all free fields $\mathbb{A}^{(j)}$
to be massive. In this case both Wick product factors of theorem \ref{WickDecomposition}, when evaluated at the test 
functions $\phi$ and, respectively, $\varphi$, are operators continuously transforming the Hida space $(E)$ into itself, and can be composed. 
In this case the contractions $\kappa'_{{}_{l',m'}}(\phi) \otimes_{k} \kappa''_{{}_{l'',m''}}(\varphi)$
have ordinary meaning given by the ordinary dual pairings. In the second step, we consider the case
in which some free fields $\mathbb{A}^{(j)}$ are massless. 
We replace $p_0(\boldsymbol{p}) = | \boldsymbol{p}|$ in the exponent $e^{\mp ip\cdot x}$ defining the plane wave kernels
$\kappa_{0,1}, \kappa_{1,0}$ of the massless free field $\mathbb{A}^{(j)}$, with the following
massive counterpart $p_0(\boldsymbol{p}) = \sqrt{| \boldsymbol{p}|^2 + \epsilon^2}$. 
Finally, we pass to the limit $\epsilon \rightarrow 0$.

\subsection{A class of integral kernel operators that allows (tensor) product operation}

We are going to show consistency of the Bogoliubov's causality axioms (I)-(V), with free fields $\mathbb{A}^{(j)}$, their pointwise Wick products and the
(tensor) products of the pointwise Wick products of free fields, understood as finite 
sums of integral kernel operators, and finally products of (tesor) products of Wick pointwise products of free fields by 
tempered translationally invariant distributions, understood as integral kernel operators. The above said operations are sufficient for the axioms as well as for the inductive construction of the scattering operator, determined by the axioms.  Then, in order to show the consistency of these axioms we should 
prove that, indeed, the said operations are meaningful.

We investigate the general class 
of generalized operators $\Xi'$, equal to finite sums of integral kernel operators
\[
\Xi' = \sum\limits_{l',m'} \Xi_{l',m'}(\kappa'_{l',m'}) \in L\big(\mathcal{E}_{{}_{'}} , L((E),(E)^*) \big), 
\,\,\, \mathcal{E}_{{}_{'}} = \mathcal{S}(\mathbb{R}^{4k'})
\]
for which the (tensor) product operation is well-defined and the Wick decomposition through the normal ordering is applicable to their products.

The problem is that in general this class necessary should include the
operators of
\[
L\big(\mathcal{E}_{{}_{'}}, L((E),(E)^*) \big),
\]
which do not belong to
\[
L\big(\mathcal{E}_{{}_{'}}, L((E),(E)) \big).
\]
Indeed the interaction Lagrangian $\mathcal{L}(x)$ is in general equal to a pointwise Wick product (\ref{PointWiseWick}) of free fields. In case all
free fields $\mathbb{A}^{(j)}$ are massive in (\ref{PointWiseWick}) the pointwise product belongs to 
\[
L\big(\mathcal{E}_{{}_{'}}, L((E),(E)) \big).
\]
But in general the Wick pointwise product $\mathcal{L}(x)$ includes massless free field factors, as is the case e.g. for QED,
with $\mathcal{L}(x)$ including the massless e.m. potential field, so that
\begin{gather*}
\mathcal{L} \in L\big(\mathcal{E}_{{}_{'}} , L((E),(E)^*) \big), \,\,\, \textrm{but} \,\,\,
\mathcal{L} \notin L\big(\mathcal{E}_{{}_{'}} , L((E),(E)) \big)
\\
\,\,\, \mathcal{E}_{{}_{'}} = \mathcal{S}(\mathbb{R}^{4})
\end{gather*}
in this case. Therefore, we proceed, as in the proof of theorem \ref{WickDecomposition}, in two steps. In the first step we replace the 
zero momentum $p_0(\boldsymbol{p}) = |\boldsymbol{p}|$ in the exponents of the kernels $\kappa_{0,1}, \kappa_{1,0}$ of all massless free fields 
$\mathbb{A}^{(j)}$ of all Wick pointwise products of free fields 
containing massless fields, by $p_0(\boldsymbol{p}) = \sqrt{| \boldsymbol{p}|^2 + \epsilon^2}$. 
In the second step, we pass to the limit $\epsilon \rightarrow 0$.

Using this method we show
\begin{theorem}
The class within which the (tensor) product of generalized operators (understood as finite sums of integral kernel operators
with vector valued kernels) is well-defined as a finite sum of integral kernel operators, includes all operators of the form   
\begin{equation}\label{ProductClassGenOp}
t(x_1, \ldots, x_n) \, {:}W_1(x_1)W_2(x_2) \ldots W(x_k){:}, \,\,\, t \in  \mathcal{S}(\mathbb{R}^{4k})^* = \mathcal{S}(\mathbb{R}^{4})^{* \, \otimes \, k},
\end{equation}
with translationally invariant $t$, and $W_i$ being Wick pointwise products of massless or massive free fields. 
\qed
\label{ClassWithProductThm}
\end{theorem}

For operators $\Xi', \Xi''$ in this class there exist $\epsilon$-approximations
\begin{gather*}
\Xi'_{{}_{\epsilon}} = \sum\limits_{l',m'} \Xi_{l,m}(\kappa'_{\epsilon \, l,m}) \in L\big(\mathcal{E}_{{}_{'}}, L((E),(E)) \big), 
\\
\Xi''_{{}_{\epsilon}} = \sum\limits_{l'',m''} \Xi_{l,m}(\kappa''_{\epsilon \, l'',m''}) 
\in L\big(\mathcal{E}_{{}_{''}}, L((E),(E)) \big), 
\end{gather*}
(here with $\mathcal{E}_{{}_{'}}$ and $\mathcal{E}_{{}_{''}}$ being some Schwartz spaces of $\mathbb{C}^{d_i}$-valued functions) for which 
\[
\Xi'_{{}_{\epsilon}} \longrightarrow \Xi',
\,\,\,
\Xi''_{{}_{\epsilon}} \longrightarrow \Xi''
\]
in 
\[
L\big(\mathcal{E}_{{}_{(i)}}, L((E),(E)^*) \big)
\]
and moreover, the Fock decomposition is naturally applicable
to their operator composition product $\Xi_{{}_{\epsilon}}$
\[
\Xi_{{}_{\epsilon}} (\phi\otimes\varphi) \overset{\textrm{df}}{=} \Xi'_{{}_{\epsilon}} (\phi) \circ \Xi''_{{}_{\epsilon}}(\varphi) 
\]
and such that the kernels
\[
\kappa'_{\epsilon, l',m'} \otimes_q \kappa''_{\epsilon \, l'',m''}
\]
of the Wick decomposition of the operator product $\Xi_{{}_{\epsilon}}$ converge, when $\epsilon \rightarrow 0$, to some kernels
\[
\kappa'_{l',m'} \otimes\big|_k \kappa''_{l'',m''} \in L(E^{\otimes(l'+l''+m'+m''-2k)}, \mathcal{E}_{{}_{'}}^{*}\otimes \mathcal{E}_{{}_{''}}^{*})
\]
(here, for simplicity of notation, we have assumed all nuclear single particle momentum test spaces equal $E$, meaning that all involved free fields
are assumed to have equal number of components) representing a well-defined finite sum $\Xi$ of generalized operators. 
%This is in particular 
%the case if the contraction integrals 
%\[
%\kappa'_{l',m'} \otimes_q \kappa''_{l'',m''}
%\]
%are convergent and represent well-defined elements of
%\[
%L(E^{\otimes(l'+l''+m'+m''-2k)}, \mathcal{E}_{{}_{'}}^{*}\otimes \mathcal{E}_{{}_{''}}^{*}).
%\]
Thus, in this class the product operator $\Xi_{{}_{\epsilon}}$ converges
\[
\Xi_{{}_{\epsilon}} \longmapsto \Xi \in L\big(\mathcal{E}_{{}_{'}}\otimes \mathcal{E}_{{}_{''}}, L((E),(E)^*) \big)
\]
to an operator $\Xi$ in 
\[
L\big(\mathcal{E}_{{}_{'}}\otimes \mathcal{E}_{{}_{''}}, L((E),(E)^*) \big).
\]
In practice, we construct the $\epsilon$-approximation within the class indicated above, just by replacing the exponents of the
free massless field plane wave kernels with exponents in which $p_0(\boldsymbol{p}) = |\boldsymbol{p}|$ is replaced by 
$p_0(\boldsymbol{p}) = (|\boldsymbol{p}|^2+\epsilon^2)^{1/2}$.

This definition of product can be generalized over a still more general finite sums of integral kernel operators
\begin{gather*}
\Xi' = \sum\limits_{l',m'} \Xi'_{l',m'}(\kappa'_{l',m'}) \in L\big(\mathcal{E}_{{}_{'}}, L((E),(E)^*) \big), 
\,\,\, \mathcal{E}_{{}_{'}} = \mathcal{S}(\mathbb{R}^{4k'})
\\
\Xi'' = \sum\limits_{l'',m''} \Xi''_{l'',m''}(\kappa''_{l'',m''}) \in L\big(\mathcal{E}_{{}_{''}}, L((E),(E)^*) \big), 
\,\,\, \mathcal{E}_{{}_{''}} = \mathcal{S}(\mathbb{R}^{4k''}),
\end{gather*}
provided only they possess the $\epsilon$-approximations 
\begin{gather*}
\Xi'_{{}_{\epsilon}} = \sum\limits_{l',m'} \Xi'_{l',m'}(\kappa'_{\epsilon \, l',m'}) \in L\big(\mathcal{E}_{{}_{'}}, L((E),(E)) \big), 
\,\,\, \mathcal{E} = \mathcal{S}(\mathbb{R}^{4k'})
\\
\Xi''_{{}_{\epsilon}} = \sum\limits_{l'',m''} \Xi_{l'',m''}(\kappa''_{\epsilon \, l'',m''}) \in L\big(\mathcal{E}_{{}_{''}}, L((E),(E)) \big), 
\,\,\, \mathcal{E} = \mathcal{S}(\mathbb{R}^{4k''}),
\end{gather*}
with the kernels 
\[
\kappa'_{\epsilon, l',m'} \otimes_q \kappa''_{\epsilon \, l'',m''}
\]
of the Wick decompositions of their products
converging in the sense defined above. In fact the class of  generalized operatos, on which product operation is well-defined, 
can still be extended over infinite sums 
of integral kernel operators of the class (\ref{ProductClassGenOp}), provided the infinite sums represent Fock expansions convergent in the sense
of \cite{obataJFA}.  In application to causal QFT, where the Lagrangian interaction density operator $\mathcal{L}(x)$
is equal to a Wick polynomial in free fields of finite degree, the class  (\ref{ProductClassGenOp}), allowing
the product operation, is sufficient.

In the first step, indicated above, we need to show that indeed in case all free fields are massive, or in case the exponents 
of the kernles of the massless fields are replaced by the exponents with $p_0(\boldsymbol{p}) = (|\boldsymbol{p}|^2+\epsilon^2)^{1/2}$,
the operators $\Xi', \Xi''$ of the said class, and their (tensor) product, defined by the composition $\Xi' \circ \Xi''$ indeed belong, respectively, to  
\[
L\big(\mathcal{E}_{{}_{'}}, \mathcal{L}((E),(E)) \big), L\big(\mathcal{E}_{{}_{''}}, \mathcal{L}((E),(E)) \big),
L\big(\mathcal{E}_{{}_{'}}\otimes \mathcal{E}_{{}_{''}}, L((E),(E)) \big)
\]
and (after being evaluated at the test functions) can be composed. In the second step, we need to show that the $\epsilon \rightarrow 0$
limit of these operators and of their composition defines a finite sum of integral kernel operators, respectively, in
\[
L\big(\mathcal{E}_{{}_{'}}, \mathcal{L}((E),(E)^*) \big), L\big(\mathcal{E}_{{}_{''}}, \mathcal{L}((E),(E)^*) \big), 
L\big(\mathcal{E}_{{}_{'}}\otimes \mathcal{E}_{{}_{''}}, L((E),(E)^*) \big).
\]
We will not present here the proof of theorem \ref{ClassWithProductThm}, 
but mention only that the proof is based on the following two theorems \ref{Hida-ObataTheorem} 
and \ref{ExtensionThm}, applied for 
$V = \mathcal{E}_{{}_{'}}, \mathcal{E}_{{}_{''}}, \mathcal{E}_{{}_{'}}\otimes \mathcal{E}_{{}_{''}}$.
\begin{theorem}[Hida, Obata\cite{obataJFA}]
\[
\Xi(\kappa_{lm}) \in
\begin{cases}
L\big(V, \, L((E),(E)^*)\big), 
\\
L\big(V, \, L((E),(E))\big),\\
\end{cases}
\]
\begin{center}
if and only if
\end{center}
\[
\begin{cases}
\kappa_{lm} \in 
L\big(E^{\widehat{\otimes} \, (l+m)}, V^*\big), 
\\
\text{$\kappa_{lm}$ can be extended to a separately cont. map}:  E^{*\widehat{\otimes} l} \times  E^{\widehat{\otimes} m} \longrightarrow V^*,\\
\end{cases}
\]
\qed
\label{Hida-ObataTheorem}
\end{theorem}

Let $E, E_1 , F, F_1$  be t.v.s. such that $E, F$ are, respectively, dense in $E_1, F_1$. 
Suppose that $\mathfrak{S}$ is a family of bounded subsets of $E$ with
the property that $\mathfrak{S}_1$ covers $E_1$ , where $\mathfrak{S}_1$ denotes the family of the closures,
taken in $E_1$ , of all subsets in $\mathfrak{S}$; analogously let $\mathfrak{F},\mathfrak{F}_1$ be such families
in  $F, F_1$; finally, let $G$ be a quasi-complete Hausdorff t.v.s. Under these assumptions,
the following extension theorem holds (compare e.g. the proposition 5.4, Chap. III.5.4 in the book\cite{Schaefer} of H. H. Schaefer):

\begin{theorem}
Every $(\mathfrak{S}, \mathfrak{F})$-hypocontinuous bilinear mapping of $E \times F$ into $G$ has a unique
extension to $E_1 \times F_1$ (and into $G$) which is bilinear and$(\mathfrak{S}_1, \mathfrak{F}_1)$-hypocontinuous.
\qed
\label{ExtensionThm}
\end{theorem}

\subsection{Bogoliubov's causality axioms}

Let $\mathcal{L}(x)$ be the interaction Lagrangian, equal to a pointwise Wick product of free fields, or
to a pointwise Wick polynomial in free fields. Let $g$ be the intensity of interaction scalar function, or more generally, we consider many-component
intensity of interaction function $(g,h)$ and a modified interaction Lagrangian $g\mathcal{L}$ or $g\mathcal{L}+h\mathbb{A}$,
for any pointwise Wick polynomial  $\mathbb{A}$ in free fields of even degree in Fermi fields.  

Let
\begin{gather*}
g^{\otimes \, n}  \in \mathcal{E}^{\otimes \, n} 
\,\,\,\,\,\,\,\,\,
\textrm{or, respectively,}
\,\,\,\,\,\,\,\,\,
g^{\otimes \, n} \otimes h^{\otimes \, k} \in \mathcal{E}^{\otimes \, n} \otimes (\oplus_{1}^{d}\mathcal{E})^{\otimes \, k}.
\end{gather*}

For such $g$ or $(g,h)$ we construct the scattering operator $S(g)$ or, more generally, $S(g,h)$, with the modified
intensity, $g$ or $(g,h)$, of interaction as the following series  
\begin{gather*}
S(g) = \boldsymbol{1} + \sum\limits_{n=1}^{\infty} {\textstyle\frac{1}{n!}} S_n(g^{\otimes \, n}),
\,\,\,\,
S(g)^{-1} = \boldsymbol{1} + \sum\limits_{n=1}^{\infty} {\textstyle\frac{1}{n!}} \overline{S_n}(g^{\otimes \, n}),
\\
\textrm{or}
\,\,\,\,
S(g,h) = \boldsymbol{1} 
+ \sum\limits_{n=1}^{\infty} \sum\limits_{p=0}^{n} {\textstyle\frac{1}{n!}} S_{n-p,p}(g^{\otimes \, (n-p)} \otimes h^{\otimes \, p})
\end{gather*}
(denoted also by $S(g\mathcal{L})$, $S(g\mathcal{L})^{-1}$ or, respectively, $S(g\mathcal{L}+h\mathbb{A})$),
with each $S_n$ or $S_{n,k}$, understood as integral kernel operators 
\begin{multline*}
S_n(g^{\otimes \, n})  = \sum_{l,m}
\int \kappa_{lm}\big(\boldsymbol{p}_1, \ldots, \boldsymbol{p}_l,
\boldsymbol{q}_1, \ldots, \boldsymbol{q}_m;  g^{\otimes \, n} \big) \,\,
\partial_{\boldsymbol{p}_1}^{*} \ldots \partial_{\boldsymbol{p}_l}^{*} 
\partial_{\boldsymbol{q}_1} \ldots \partial_{\boldsymbol{q}_m} \,\, \times
\\
\times \,
d\boldsymbol{p}_1 \ldots d\boldsymbol{p}_l
d\boldsymbol{q}_1 \ldots d\boldsymbol{q}_m
\\
= 
\int d^4 x_1 \ldots d^4x_n \, S_n(x_1, \ldots, x_n) \, g(x_1) \ldots g(x_n),
\end{multline*}
or
\begin{multline*}
S_{n,k}(g^{\otimes \, n} \otimes h^{\otimes \, k})  = \sum_{l,m}
\int \kappa_{lm}\big(\boldsymbol{p}_1, \ldots, \boldsymbol{p}_l,
\boldsymbol{q}_1, \ldots, \boldsymbol{q}_m;  g^{\otimes \, n} \otimes h^{\otimes \, k} \big) \,\, \times
\\
\times \,
\partial_{\boldsymbol{p}_1}^{*} \ldots \partial_{\boldsymbol{p}_l}^{*} 
\partial_{\boldsymbol{q}_1} \ldots \partial_{\boldsymbol{q}_m} 
d\boldsymbol{p}_1 \ldots d\boldsymbol{p}_l
d\boldsymbol{q}_1 \ldots d\boldsymbol{q}_m
\\
= 
\int d^4 x_1 \ldots d^4x_n d^4 y_1 \ldots d^4y_k \, S_{n,k}(x_1, \ldots, x_n, y_1, \ldots, y_k) \, g(x_1) \ldots g(x_n),
\end{multline*}
with vector-valued distributional kernels $\kappa_{lm}$ in the sense of Obata\cite{obataJFA}, with the values in the distributions
$V^*$ over the test nuclear space
\begin{gather*}
V=\mathcal{E}^{\otimes \, n} \ni g^{\otimes \, n} 
\,\,\,\,\,\,\,\,\,
\textrm{or, respectively,}
\,\,\,\,\,\,\,\,\,
V=\mathcal{E}^{\otimes \, n}\otimes (\oplus_{1}^{d}\mathcal{E})^{\otimes \, k}  \ni g^{\otimes \, n} \otimes h^{\otimes \, k}.
\end{gather*}

Let us recall Bogoliubov's axioms for $S$. We confine attention to the simpler, but essential, case $S(g)$ with only the scalar
intensity of interaction function $g$. The more general case $S(g,h)$ is analogous and brings no essentially new difficulties
into the whole analysis.
 
Bogoliubov's causality axioms (I)-(V) for $S(g)$ read
\begin{enumerate}
\item[(I)]
\[
              S(g_1+g_2) = S(g_2)S(g_1), \,\,\, \textrm{whenever} \,\, \textrm{supp} \, g_1 \preceq \textrm{supp} \, g_2. 
\]

\item[(II)]
\[
              U_{a,\Lambda}S(g)U_{b, \Lambda}^{+} = S(T_{{}_{b,\Lambda}}g), \,\,\, 
T_{{}_{b,\Lambda}}g(x) = g(\Lambda x +b). 
\]

\item[(III)]
\[
\eta S(g)^+\eta = S(g)^{-1}.   
\]

\item[(IV)]
\[               
S_1(x_1) = i \mathcal{L}(x_1)
\]
where $\mathcal{L}(x_1)$ is the interaction Lagrangian density operator.

\item[(V)] 
The value of the retarded part of a vector valued kernel should coincide with the natural formula given by the multiplication by the step theta 
function on a space-time test function, whenever the natural formula is meaningful for this test function.
\end{enumerate}

The axiom (V) has been added by Epstein and Glaser\cite{Epstein-Glaser}. These axioms should be understood in the order by order sense, and
as such, they can be rewritten in the following manner
\begin{enumerate}
\item[(I)]
\begin{gather*}
              S_{n}(x_1, \ldots, x_{n}) = S_{k}(x_1, \ldots, x_{k})S_{n-k}(x_{k+1}, \ldots, x_{n}),
\\
\,\,\,\, \textrm{whenever $\{x_{k+1}, \ldots, x_{n} \} \preceq \{x_1, \ldots, x_k\}$},  % \\
\end{gather*}

\item[(II)]
\[
U_{b,\Lambda} S_n(x_1, ..,x_n) U_{b, \Lambda}^{+} = S_n(\Lambda^{-1}x_1 - b, .., \Lambda^{-1}x_n - b),
\]

\item[(III)]
\[     
\overline{S}_{n}(x_1, \ldots, x_{n}) = \eta S_n(x_1, \ldots, x_{n})^{+} \eta,
\]

\item[(IV)]
\[                  
S_{1}(x_1) = i \mathcal{L}(x_1), 
\]

\item[(V)]
The singularity degree of the retarded part of a kernel
should coincide with the singularity degree of this kernel, for the kernels of the generalized integral kernel 
and causal operators $D_{(n)}$ which are equal to linear combinations of products of the generalized operators $S_k$.
\end{enumerate}

We are going tho show that (I)-(V) are meaningful whenever $S_n$ are understood as integral kernel operators. 
In order to do it it is sufficient to show that the (tensor) products 
\[
S_{k}(x_1, \ldots, x_{k})S_{n-k}(x_{k+1}, \ldots, x_{n})
\]
are meaningful as integral kernel operators. Because we start with $S_1 = \mathcal{L}$ which is equal to a pointwise Wick product of free fields (or pointwise Wick
polynomial in free fields), and beasue the Epstein-Glaser inductive construction of $S_n$ uses only the mentioned (tensor) product,
then indeed the consistency will follow from theorem \ref{ClassWithProductThm}. As we will see, if we start from the 
pointwise Wick polynomial $S_1= \mathcal{L}$ of free fields, then the inductive step will give operators $S_n$ of the class
of theorem \ref{ClassWithProductThm}.

\subsection{Inductive step}\label{InductiveStep}

Let us recall the inductive step construction for $S_n$ (the case of $S_{n.k}$ is analogous).
Having given $\{S_k\}_{{}_{k\leq n-1}}$ 
we want to construct $S_n(Z, x_n) = S(Z, x_n)$. Here $X \cup Y = \{x_1, \ldots, x_n\} = Z$, $X \cap Y = \emptyset$)
denote the disjoint subsets of the set $Z$ of space-time variables $x_1, \ldots, x_n$. 
To this end we construct, after Epstein and Glaser\cite{Epstein-Glaser} or Scharf\cite{Scharf}, 
the following generalized operators
\[
\begin{split}
A'_{(n)}(Z, x_n)  
= \sum \limits_{X\sqcup Y=Z, X\neq \emptyset} \overline{S}(X)S(Y,x_n), \,\,\,
\\
R'_{(n)}(Z, x_n)  
= \sum \limits_{X\sqcup Y=Z, X\neq \emptyset} S(Y,x_n)\overline{S}(X),
\end{split}
\]
and then
\[
\begin{split}
A_{(n)}(Z, x_n) = \sum \limits_{X\sqcup Y=Z} \overline{S}(X)S(Y,x_n)
= A'_{(n)}(Z, x_n)  + S(x_1, \ldots, x_n), \\
R_{(n)}(Z, x_n) = \sum \limits_{X\sqcup Y=Z} S(Y,x_n)\overline{S}(X)
= R'_{(n)}(Z, x_n) + S(x_1, \ldots, x_n),
\end{split}
\]
From (I)-(V) it follows that $A_{(n)}$ and $R_{(n)}$ have causal supports in space-time variables
$x_1, \ldots, x_n$ and that  $A_{(n)}$ is an advanced and $R_{(n)}$ a retarded generalized operator,
compare \cite{Epstein-Glaser} or \cite{Scharf}. 
Thi means that the support of $R_{(n)}$, resp. $A_{(n)}$, in each variable  $x_1, \ldots, x_{n-1}$
is contained within the forward, resp. past, light cone emerging from $x_n$.
These proofs\cite{Epstein-Glaser, Scharf} were performed for $S_k$ understood
as Wightman operator distributions, but remain identical for $S_k$ understood as integral kernel
operators, whenever $S_k$ are well-defined as integral kernel operators. But that $S_k$ are indeed well-defined as integral
kernel operators can be seen inductively on application of theorem \ref{ClassWithProductThm}.

Therefore
\[
D_{(n)} = R'_{(n)} - A'_{(n)} = R_{(n)} - A_{(n)} 
\,\,\,\,\,\,\,\,\,\,\,\,\,\,\,\, \textrm{is causally supported}
\]
and
\[
R_{(n)} \,\,\, \textrm{-- is a retarded part of $D_{(n)}$}
\,\,\,\,\,\,\,\,\,\,\,\,\,\,\,\,\,\,\,\,
A_{(n)} \,\,\, \textrm{-- is an advanced part of $D_{(n)}$}
\]
so that
\[
\boxed{
 S(x_1, \ldots, x_n) =  R_{(n)}(x_1, \ldots, x_n)-R'_{(n)}(x_1, \ldots, x_n) 
}
\]
can be computed, as the splitting of causally supported generalized operator into the retarded and advanced part
can be computed independently of the axioms (I)-(V). In practice, we apply Wick decomposition to the generalized operator
$D_{(n)}$. All scalar factors multiplying the Wick monomials in this decomposition are causally supported tempered distributions.
Thus, computation is reduced to the computation of the splitting of causal scalar tempered distributions into the
retarded and advanced parts. As is well-known, such splitting is non-unique if the singularity degree at zero
of the splitted distribution is equal or greater than zero (in space-time variables). This reduces the whole problem
to the computation of the ultraviolet quasi-asymptotics of the splitted distributions. For the interaction Lagrangians 
$\mathcal{L}$ equal to pointwise Wick polynomials in free fields, the ultraviolet (in space-time variables) 
asymptotics of the causal distributions turns out to be trivial, as the Fourier transforms of the scalar causal 
distributions which are essential (corresponding to the loop-like divergent Feymnam diagrams) are regular 
function-like distributions, which behave polynomially at
infinity. The degree of these asymptotic polynomials gives us the singularity degree of the splitted causal distributions,
so that the full theory of quasi-asymptotics of tempered distributions is in fact not needed here.

\section{Scattering operator and interacting fields}

Application of the  Hida operators as above converts the free fields, 
their Wick products $\mathbb{A}$ and the $n$-th order contributions 
\[
S_n(g^{\otimes \, n}) \,\, 
\,\,\,\,\,\,\,\,\,\,\,\,\,\,\,\,\,\,\,\,\,
\textrm{and}
\,\,\,\,\,\,\,\,\,\,\,\,\,\,\,\,\,\,\,\,\,
\mathbb{A}_{{}_{\textrm{int}}}^{(n)}(g^{\otimes \, n},\phi) 
\]
written frequently as
\[
S_n(g) \,\,
\,\,\,\,\,\,\,\,\,\,\,\,\,\,\,\,\,\,\,\,\,
\textrm{and}
\,\,\,\,\,\,\,\,\,\,\,\,\,\,\,\,\,\,\,\,\,
\mathbb{A}_{{}_{\textrm{int}}}^{(n)}(g,\phi) ,
\]
to the scattering operator
\begin{gather*}
S(g) = \boldsymbol{1} + \sum\limits_{n=1}^{\infty} {\textstyle\frac{1}{n!}} S_n(g^{\otimes \, n}),
\,\,\,\,
S(g)^{-1} = \boldsymbol{1} + \sum\limits_{n=1}^{\infty} {\textstyle\frac{1}{n!}} \overline{S_n}(g^{\otimes \, n}),
\\
\textrm{or}
\,\,\,\,
S(g,h) = \boldsymbol{1} 
+ \sum\limits_{n=1}^{\infty} \sum\limits_{p=0}^{n} {\textstyle\frac{1}{n!}} S_{n-p,p}(g^{\otimes \, (n-p)} \otimes h^{\otimes \, p})
\end{gather*}
(denoted also by $S(g\mathcal{L})$, $S(g\mathcal{L})^{-1}$ or, respectively, $S(g\mathcal{L}+h\mathbb{A})$) and to the interacting fields 
\[
\mathbb{A}_{{}_{\textrm{int}}}(g,\phi) = \int{\textstyle\frac{i\delta}{\delta h(x)}}S(g\mathcal{L}+h\mathbb{A})^{-1}S(g\mathcal{L})\Big|_{{}_{h=0}} \phi(x) dx,
\]
into the finite sums of generalized integral kernel operators
\begin{multline*}
\Xi(\kappa_{lm}) = 
\int \kappa_{lm}\big(\boldsymbol{p}_1, \ldots, \boldsymbol{p}_l,
\boldsymbol{q}_1, \ldots, \boldsymbol{q}_m\big) \,\,
\partial_{\boldsymbol{p}_1}^{*} \ldots \partial_{\boldsymbol{p}_l}^{*} 
\partial_{\boldsymbol{q}_1} \ldots \partial_{\boldsymbol{q}_m} 
\\
\times \,
d\boldsymbol{p}_1 \ldots d\boldsymbol{p}_l \, \times
d\boldsymbol{q}_1 \ldots d\boldsymbol{q}_m,
\end{multline*}
e.g. for the contributions $S_n$:
\begin{multline*}
S_n(g^{\otimes \, n})  = \sum_{l,m}
\int \kappa_{lm}\big(\boldsymbol{p}_1, \ldots, \boldsymbol{p}_l,
\boldsymbol{q}_1, \ldots, \boldsymbol{q}_m;  g^{\otimes \, n} \big) \,\,
\partial_{\boldsymbol{p}_1}^{*} \ldots \partial_{\boldsymbol{p}_l}^{*} 
\partial_{\boldsymbol{q}_1} \ldots \partial_{\boldsymbol{q}_m}
\\
\times \, 
d\boldsymbol{p}_1 \ldots d\boldsymbol{p}_l \, \times
d\boldsymbol{q}_1 \ldots d\boldsymbol{q}_m
\,\, = \,\,
\int d^4 x_1 \ldots d^4x_n \, S_n(x_1, \ldots, x_n) \, g(x_1) \ldots g(x_n),
\end{multline*}
with vector-valued distributional kernels $\kappa_{lm}$ in the sense of Obata\cite{obataJFA}, with the values in the distributions
$V^*$ over the test nuclear space
\begin{gather*}
V=\mathcal{E} \ni \phi,
\,\,\,\,\,\,\,\,\,
\textrm{or}
\,\,\,\,\,\,\,\,\,
V=\mathcal{E}^{\otimes \, (n-p)}\otimes (\oplus_{1}^{d}\mathcal{E})^{\otimes \, p}  \ni g^{\otimes \, (n-p)} \otimes h^{\otimes \, p},
\\
\textrm{or}
\,\,\,\,\,
V=\mathcal{E}^{\otimes \, n} \ni g^{\otimes \, n} 
\,\,\,\,\,\,\,\,\,
\textrm{or, respectively,}
\,\,\,\,\,\,\,\,\,
V=\mathcal{E}^{\otimes \, n} \otimes (\oplus_{1}^{d}\mathcal{E}) \ni g^{\otimes \, n} \otimes \phi
\end{gather*}
with 
\[
\mathcal{E} = \mathcal{S}(\mathbb{R}^4;\mathbb{C}). 
\]
Each of the $3$-dim Euclidean integration $d\boldsymbol{p}_i$ with respect to the spatial momenta $\boldsymbol{p}_i$ components
$\boldsymbol{p}_{i1}, \boldsymbol{p}_{i2}, \boldsymbol{p}_{i3}$,
also includes here summation over the corresponding discrete spin components $s_i\in(1,2,\ldots)$ hidden under the symbol $\boldsymbol{p}_i$.

The class to which the operators $S_n$ and $\mathbb{A}_{{}_{\textrm{int}}}^{(n)}$ belong, expressed in terms of the Hida test space,
depend on the fact if there are massless free fields present in the interaction Lagrange density operator $\mathcal{L}$ or not.
Namely: 
\begin{theorem}
\[
S_n \in
\begin{cases}
L\big(\mathcal{E}^{\otimes \, n}, \, L((E),(E))\big), 
& \text{if all fields in $\mathcal{L}$ are massive},\\
L\big(\mathcal{E}^{\otimes \, n}, \, L((E),(E)^*)\big), & \text{if there are massless fields in $\mathcal{L}$}.\\
\end{cases}
\]
\qed
\end{theorem}
and the same holds for $S_{n,p}$ with $\mathcal{E}^{\otimes \, n}$ repaced by 
$\mathcal{E}^{\otimes \, n} \otimes (\oplus_{1}^{d}\mathcal{E})^{\otimes \, p}$ 
if both $g$ and $h$, are, respectively, $\mathbb{C}$ and $\mathbb{C}^d$-valued Schwartz test functions. 
But the same distribution valued kernels
of $S_{n,p}$ can be evaluated at the \emph{Grassmann-valued test functions} $h$, 
in the sense of Berezin\cite{Berezin}, which are used in case we have the modified Lagrangian
$g\mathcal{L}+h\mathbb{A}$,
with any pointwise Wick polynomial  $\mathbb{A}$ in free fields which is of odd degree in Fermi fields. 
So that in this case we have
\begin{theorem}
\[
S_{n,p} \in
\begin{cases}
L\big((E), (E)\big) \otimes L(\mathcal{E}^{\otimes \, n} \otimes \mathcal{E}^p, \mathcal{F}^{p \, *}), 
& \text{if all fields in $\mathcal{L}$ are massive},\\
L\big((E), (E)^*\big) \otimes L(\mathcal{E}^{\otimes \, n} \otimes \mathcal{E}^p, \mathcal{F}^{p \, *}), 
& \text{if there are massless fields in $\mathcal{L}$},\\
\end{cases}
\]
with $\mathcal{F}^{p \, *}$ being the subspace of grade $p$ of the \emph{abstract Grassmann algebra} $\oplus_p\mathcal{F}^{p \, *}$ 
\emph{with inner product and involution}
in the sense of Berezin. $\mathcal{F}^p$ denotes the space of Grassmann-valued test functions $h^p$ of grade $p$ 
due to Berezin, and replacing ordinary test functions $h^{\otimes \, p}$.
\qed
\end{theorem}
Recall that $L(E_1,E_2)$ denotes the linear space of linear continuous operators $E_1\longrightarrow E_2$ endowed with the natural topology of uniform
convergence on bounded sets.

\subsection{Adiabatic limit for interacting fields}\label{IntFields}

Using Hida operators and integral kernel operators as above, we are able to solve positively the 
\emph{adiabatic limit problem}. In QED the limit $g \rightarrow 1$ of the $n$-th order contributions
$A^{(n)}_{{}_{\textrm{int}}}(g)$, $\boldsymbol{\psi}^{n}_{{}_{\textrm{int}}}(g)$ to interacting e.m. 
potential and the charged massive fields, exists and equal to a finite sum of integral kernel operators with $\oplus_{1}^{4} \mathcal{E}^*$-valued kernels
in the sense of Obata\cite{obataJFA}, and belongs in general to
\[
L\big( \oplus_{1}^{4} \mathcal{E}, \, L((E), (E)^*) \big).
\]
This limit exists if and only if the normalization in the splitting of the causal scalar tempered distributions 
into retarded and advanced parts in the computation of the higher order contributions $S_n$ and $S_{n,k}$ 
to the scattering operator is  ``natural''. Moreover, this limit exists if and only if the charged field is massive. 
\emph{This result can be interpreted as a theoretical proof of the experimentally observed fact that all electrically
charged particles are massive.}  

We do not present the complete proof but give only an example of higher order contributions which illustrates 
the role of the choice of the ``natural'' normalization in the splitting for the existence of this limit and 
which illustrates why the charged field should necessarily be massive. In the next subsection
we will try to explain the essential role of the Hida operators for the existence of the limit at all and why this limit does not exist
in general (with massive or massless charged field) if we are using operator valued distributions in the Wightman
sense instead of the generalized integral kernel operators with vector valued kernels.

Let $D_{0}$ be the Pauli-Jordan distribution of the massless scalar field and let $D_{0}^{{}^{\textrm{av}}}$ 
be its advanced part. Let $\Pi_{\mu\nu}(x_1-x_2)$ be the vacuum polarization distribution -- the coefficient in the second order
contribution $S_2(x_1,x_2)$, multiplying the (tensor) Wick product ${:}A_\mu(x_1)A_{\nu}(x_2) {:}$ -- computed
with the help of the splitting into advanced and retarded part in accordance to the inductive step referred to in
subsection \ref{InductiveStep}, with the corresponding Fourier transforms of $\Pi_{\mu\nu}$ and $\Pi^{{}^{\textrm{av}}}_{\mu\nu}$
(`natural'' normalization in the splitting is assumed)
\begin{multline}\label{Pi}
\widetilde{{\Pi}_{\mu \nu}}(p) = 
(2\pi)^{-4} \big({\textstyle\frac{p_\mu p_\nu}{p^2}} - g_{\mu\nu}\big) \widetilde{\Pi}(p),
\\
\widetilde{\Pi}(p) =
{\textstyle\frac{1}{3}} p^4
\int\limits_{4m^2}^{\infty} {\textstyle\frac{s+2m^2}{s^2(p^2-s+i0)}}\sqrt{1-{\textstyle\frac{4m^2}{s}}} ds,
\end{multline}
\begin{multline}\label{Piav}
\widetilde{{\Pi^{{}^{\textrm{av}}}}_{\mu \nu}}(p) = 
(2\pi)^{-4} \big({\textstyle\frac{p_\mu p_\nu}{p^2}} - g_{\mu\nu}\big) \widetilde{\Pi{{}^{\textrm{av}}}}(p),
\\
\widetilde{\Pi{{}^{\textrm{av}}}}(p) =
{\textstyle\frac{1}{3}} p^4
\int\limits_{4m^2}^{\infty} {\textstyle\frac{s+2m^2}{s^2(p^2-s- \, i \,  p_0 \, 0)}}\sqrt{1-{\textstyle\frac{4m^2}{s}}} ds.
\end{multline}
The ``natural'' normalization 
\begin{equation}\label{NaturalNormalization}
{\textstyle\frac{\widetilde{\Pi}(p)}{p^2}}\Big|_{{}_{p^2=0}} = 0,  \,\,\, \widetilde{\Pi}(0) = 0,
\end{equation}
in the splitting 
is assumed in (\ref{Pi}) and (\ref{Piav}.
Let us consider the $n=2k+1$-order sub-contribution to interacting e.m. potential field $A_{{}_{\textrm{int}} \nu}(g=1)$ in the limit $g \rightarrow 1$
containing $k$ vacuum polarization graph insertions $\Pi$ in the spinor QED. It can be written as the following repeated convolution
\begin{equation}\label{D0*Pi*...*D0*Pi*Do:psigammapsi:}
D_{0}^{{}^{\textrm{av}}} \ast \Pi^{{}^{\textrm{av}} \, \mu_k}_{\mu} \ast \ldots \ast D_{0}^{{}^{\textrm{av}}} \ast \Pi^{{}^{\textrm{av}} \, \nu}_{\mu_1} \ast
D_{0}^{{}^{\textrm{av}}}  \ast {:}\boldsymbol{\psi}^\sharp \gamma_\nu \boldsymbol{\psi}{:},
\end{equation}

Let $E_1$, $E_{2}$ be the single particle test spaces of the free fields $\boldsymbol{\psi}^\sharp, \boldsymbol{\psi}$ 
and let $\oplus_{1}^{4} \mathcal{E}$ be the space-time test space.
Let $\kappa_{1,0}^{\sharp}$, $\kappa_{0,1}^{\sharp}$ be kernels of the Dirac conjugated field  $\boldsymbol{\psi}^\sharp$,
and $\kappa_{1,0}$, $\kappa_{0,1}$ be kernels of the Dirac field  $\boldsymbol{\psi}$. 
The evaluations $\langle \kappa_{lm}(\xi_1 \otimes \xi_2), \rangle \phi$ of the kernels $\kappa_{lm}$ of the sub-contributions
(\ref{D0*Pi*...*D0*Pi*Do:psigammapsi:}) at the test functions $\xi_1 \otimes \xi_2 \otimes \phi$ $\in E_1 \otimes E_{2} \otimes (\oplus_{1}^{4} \mathcal{E})$
are equal to the $\epsilon \rightarrow 0$ limits of the integrals
\begin{multline}\label{epsilon-kernels}
\Big\langle D_{0 \, \epsilon}^{{}^{\textrm{av}}} \ast \Pi^{{}^{\textrm{av}} \, \mu_k}_{\mu} \ast \ldots \ast D_{0 \, \epsilon}^{{}^{\textrm{av}}} \ast \Pi^{{}^{\textrm{av}} \, \mu_1}_{\nu} \ast D_{0 \, \epsilon}^{{}^{\textrm{av}}}  \ast \big[\kappa_{l_1,m_1}^{\sharp}(\xi_1) \gamma^\nu \dot{\otimes} \kappa_{l_2,m_2}(\xi_2)\big], \, \phi \Big\rangle
\\
=
\sum\limits_{s_1,s_2} \int d^3 \boldsymbol{p}_1 d^3 \boldsymbol{p}_2
{\textstyle\frac{\xi_1(s_1, \boldsymbol{p}_1)\xi_2(s_2, \boldsymbol{p}_2) u_{s_1}^{\pm}(\boldsymbol{p}_1)^\sharp \gamma^\nu u_{s_2}^{\mp}(\boldsymbol{p}_2)}{[(\pm p_1\pm p_2)^2 +i \epsilon \, (\pm p_{10} \pm p_{20})]^{k+1}}} \,\, \times
\\
\times \,\,
\widetilde{\Pi^{{}^{\textrm{av}} \, \mu_k}_{\mu}}(\pm p_1 \pm p_2) \widetilde{\Pi^{{}^{\textrm{av}} \, \mu_{k-1}}_{\mu_k}}(\pm p_1 \pm p_2)
\ldots \widetilde{\Pi^{{}^{\textrm{av}} \, \mu_{1}}_{\nu}}(\pm p_1+ \pm p_2)
\widetilde{\phi}(\pm p_1\pm p_2)
\end{multline}
\[
p_1, p_2 \in \mathcal{O}_{{}_{m,0,0,0}} = \{p: \,\,\, p\cdot p = m^2, \, p_0>0  \}.
\]
Recall that $E_1=E_2 = \mathcal{S}(\mathbb{R}^3)$.
$u_{s}^{+}(\boldsymbol{p}) = u_s(\boldsymbol{p})$, $u_{s}^{-}(\boldsymbol{p}) = v_s(\boldsymbol{p})$ 
are the Fourier transforms of the basic solutions of the free Dirac equation 
%given in Appendix \ref{fundamental,u,v}
, $\gamma^\mu$ are the Dirac gamma matrices, 
%(\ref{chiralgamma})
and finally
$u_{s}^{\pm}(\boldsymbol{p})^\sharp$ is the Dirac conjugation of the spinor $u_{s}^{\pm}(\boldsymbol{p})$. The plus sign stands everywhere
in $\pm p_1$ and in $u_{s_1}^{\pm}(\boldsymbol{p}_1)^\sharp$ whenewer $(l_1,m_1) = (1,0)$. The minus
sign stands everywhere
in $\pm p_1$ and in $u_{s_1}^{\pm}(\boldsymbol{p}_1)^\sharp$ whenewer $(l_1,m_1) = (0,1)$. Analogously,
the plus sign stands everywhere
in $\pm p_2$ and minus sign in $u_{s_2}^{\pm}(\boldsymbol{p}_2)$ whenewer $(l_2,m_2) = (1,0)$. The minus
sign stands everywhere
in $\pm p_2$ and plus sign in $u_{s_2}^{\pm}(\boldsymbol{p}_2)$ whenewer $(l_2,m_2) = (0,1)$.

For the ``natural'' normalization in the Epstein-Glaser splitting, and in case $m\neq 0$, the singularity appearing in the limit
\[
{\textstyle\frac{1}{[p^2+\epsilon p_0]^{k+1}}} \overset{\epsilon\rightarrow 0}{\longrightarrow} {\textstyle\frac{1}{(p^{2})^{k+1}}}
-  \textrm{sgn} \, (p_0) \, {\textstyle\frac{i\pi(-1)^{k}}{k!}}\delta^{(k)}(p^2),
\]
is cancelled by the Fourier transform $\widetilde{\Pi^{\textrm{av} \, \mu\nu}}$ of $\Pi^{\textrm{av} \, \mu\nu}$, as $\widetilde{\Pi^{\textrm{av} \, \mu\nu}}
= (\tfrac{p^\mu p^\nu}{p^2} -g^{\mu\nu}) \widetilde{\Pi}(p)$ with a regular $\widetilde{\Pi}$ in the vicinity of the cone $p^2=0$, and equal there to 
$\widetilde{\Pi}(p) = [p^2]^2g_0(p)$ with still regular $g_0$ there. Now the freedom in normalization consists here (for Fourier transformed
$\Pi^{\mu\nu}$) in addition of a polynomial of second degree in $p$, as the singularity degree at zero of $\widetilde{\Pi}^{\mu\nu}$ is equal to two. In  particular
we can add a constant term $g^{\mu\nu}$ in (\ref{Pi}) and (\ref{Piav}), but this modification will destroy the cancellation of th singularities
so that the above $\epsilon \rightarrow 0$ in (\ref{epsilon-kernels}) will no longer exist. 
Still, in principle (freedom in the splitting),
we can add to $\widetilde{\Pi^{\mu\nu}}(p)$, or to $\widetilde{\Pi^{\textrm{av} \, \mu\nu}}(p)$ the term
of the form $f(p^2) g^{\mu\nu}$ with  $f$ which has zero of at least second order at zero. But the
kernels of some even order sub contributions
\[
\ldots \underbrace{\big(S_{{}_{\textrm{ret}, \textrm{av}}} \ast \Sigma_{{}_{\textrm{ret}, \textrm{av}}} \ast\big)}_\textrm{$k$ terms} \ldots  \ast \boldsymbol{\psi}, \,\,\, \textrm{and respectively} \,\,\,
\ldots \underbrace{\big(D_{0}^{{}^{\textrm{av}, \textrm{ret}}} \ast \Pi^{{}^{\textrm{av}, \textrm{ret}} \, \nu_k}_{\mu_k} \ast \big)}_\textrm{$k$ terms} \ldots \ast A,
\]
to interacting fields $\boldsymbol{\psi}_{{}_{\textrm{int}}}$
and, respectively, $A_{{}_{\textrm{int}}}$  are well-defined only with the stronger condition (\ref{NaturalNormalization})
put on $\Pi$.

Because the Fourier transform of the vacuum polarization in QED
with massless charged field is not smooth at the cone $p^2=0$, having the jump $\theta(p^2)$ there, and this singularity
cannot be repaired by any choice of the
splitting (addition of any polynomial in momenta of second degree), then we are confronted with the valuation of the
distribution
\[
{\textstyle\frac{1}{[\upsilon + i \epsilon]^{k+1}}}
\overset{\epsilon\rightarrow 0}{\longrightarrow}
{\textstyle\frac{1}{\upsilon^{k+1}}}
-
{\textstyle\frac{i\pi(-1)^{k}}{k!}}\delta^{(k)}(\upsilon)
\]
in single real variable $\upsilon =(p_1\pm p_2)^2$ at the ``test function'' which has the jump-type and $\sim \tfrac{1}{\sqrt{\upsilon}}$-type
singularity at $\upsilon=(p_1\pm p_2)^2=0$, which, as we know from the distribution theory,
is not well-defined, or alternatively: there is no sensible way of definition of the product of the theta function $\theta(\upsilon)$-distribution
(or the $\tfrac{1}{\sqrt{\upsilon}}$-function-type-distribution)
and the derivatives of the Dirac delta distribution $\delta^{(k)}(\upsilon)$.

By the existence of the $\epsilon \rightarrow 0$ limits in (\ref{epsilon-kernels})
defining the vector valued kernels of (\ref{D0*Pi*...*D0*Pi*Do:psigammapsi:}), 
and theorem \ref{Hida-ObataTheorem} (theorems 3.6 and 3.9  of  \cite{obataJFA} or their generalization to the Fermi case
or general Fock space) we obtain existence of the sub-contributions (\ref{D0*Pi*...*D0*Pi*Do:psigammapsi:}) 
as integral kernel operators in the adiabatic limit. But, as we have seen the $\epsilon \rightarrow 0$ limits, 
defining the kernels of (\ref{D0*Pi*...*D0*Pi*Do:psigammapsi:}) in the adiabatic limit for spinor QED with massless Dirac field,
do not exist. In general, we have

\begin{theorem}
For QED with massive charged field, the higher order contributions $\psi_{{}_{\textrm{int}}}^{(n)}(g^{\otimes \, n})$ 
and $A_{{}_{\textrm{int}}}^{(n)}(g^{\otimes \, n})$
 to interacting fields $\psi_{{}_{\textrm{int}}}$ and $A_{{}_{\textrm{int}}}$ 
in the adiabatic limit $g\rightarrow 1$ are well-defined as sums of generalized integral kernel operators
with vector valued kernels in the sense of Obata\cite{obataJFA},
\[
\underset{g\rightarrow 1}{\textrm{lim}} \psi_{{}_{\textrm{int}}}^{(n)}(g^{\otimes \, n}),
\underset{g\rightarrow 1}{\textrm{lim}} A_{{}_{\textrm{int}}}^{(n)}(g^{\otimes \, n})
\in L\big( \oplus_{1}^{d}\mathcal{E}, \, L((E),(E)^*)\big), 
\] 
and this is the case only for the ``natural''
choice in the Epstein-Glaser splitting in the construction of the scattering operator.
\label{g->1IntFields}
\end{theorem}

\vspace{0.5cm}

But:

\vspace{0.5cm}

\begin{theorem}
For causal perturbative QED on the Minkowski space-time with the Hida operators as 
the creation-annihilation operators and with massless charged field, the higher order contributions to
interacting fields  in the adiabatic limit $g\rightarrow 1$ are not well-defined, 
even as sums of generalized integral kernel operators in the sense of Obata, and for
no choice in the Epstein-Glaser splitting in the construction of the scattering operator.
\label{g->1IntFieldsm=0}
\end{theorem}

\subsection{Comparison with the approach based on Wightman's operator distributions}\label{Comparison}

In the adiabatic limit $g \rightarrow 1$ the higher order contributions 
$\mathbb{A}_{{}_{\textrm{int}}}^{(n)}(g=1)$ to interacting fields  $\mathbb{A}_{{}_{\textrm{int}}}$ in QED preserve 
in general the meaning of the generalized operators -- finite sums of integral kernel operators, \emph{i.e.} continuous maps
\[
\oplus_{1}^{4} \mathcal{E} \ni \phi \xrightarrow{\textrm{continously}} \mathbb{A}_{{}_{\textrm{int}}}^{(n)}(g=1, \phi)
\in L\big((E), (E)^*\big) \,\,\,\,\,\,\,\,\,\,\,\, \textrm{case A)}
\] 
and only for some exceptional contributions $\mathbb{A}_{{}_{\textrm{int}}}^{(n)}(g=1)$, or some of their sub-contributions, we have
\[
\oplus_{1}^{4} \mathcal{E} \ni \phi \xrightarrow{\textrm{continously}} \mathbb{A}_{{}_{\textrm{int}}}^{(n)}(g=1, \phi)
\in L\big((E), (E)\big). \,\,\,\,\,\,\,\,\,\,\,\, \textrm{case B)}
\] 
Only in case B) the contribution $\mathbb{A}_{{}_{\textrm{int}}}^{(n)}(g=1)$ can be understood as operator valued distributions
also in the Wightman sense \cite{wig}. Contributions of class A) which are not of class B) cannot be understood as operator valued distributions, 
and in particular cannot be accounted for within the approach based on operator valued distributions in the Wightman sense.
Let us recall that the Hida space $(E)$ contains the so-called fundamental domain $\mathcal{D}_0$
used in \cite{wig}, which consists of all images of the vacuum state under the polynomial expressions in 
$\boldsymbol{\psi}(f_1), \boldsymbol{\psi}^\sharp(f_2), A(f_3)$, $\ldots$, $\boldsymbol{\psi}(f_{n-1}), \boldsymbol{\psi}^\sharp(f_{n-1}), A(f_n)$,
for $n\in  \mathbb{N}$ and $f_k$ ranging over the Schwartz test functions (if we restrict the arguments $f_k$ of $A$
to the subspace $\mathcal{S}^{00}$ of the Schwartz space). Contributions of class A) which are not of class B) transform some of the Fock states
in $(E)$ into nonnormalizable states which do not belong to the Fock space, but only to the space $(E)^*$ dual to the Hida space $(E)$.
In general, the states of $\mathcal{D}_0 \subset (E)$ are transformed by the contributions of class A) into nonnormalizable states of $(E)^*$
and cannot represent any operator valued distributions in the Wightman sense. An example of the contribution
of type B) is the first order contribution $A_{{}_{\textrm{int}}}^{(1)}(g=1)$ to the interacting e.m. potential field. 

For example, the first order 
contribution $\psi_{{}_{\textrm{int}}}^{(1)}(g=1)$
in the adiabatic limit belongs to class A) but not to class B), and cannot be subsumed within the approach based on  Wightman operator
distributions. Let $E^{\pm}_1 = \mathcal{S}(\mathbb{R}^{3\times 2})$ be the positive/negative energy single particle test space
of the free Dirac field, $E_1= E^{+}_1 \oplus E^{-}_1$ -- the total single particle test space of the Dirac field, $E_2
= \mathcal{S}^{0}(\mathbb{R}^{3\times 4})$ the single particle test space of the free e.m. potential field. 
$\psi_{{}_{\textrm{int}}}^{(1)}(g=1)$ is a finite sum of well-defined integral kernel operators 
of class A), which evaluated at a test function $\phi \in \oplus_{1}^{4} \mathcal{E}$ is equal to 
\begin{align*}
\psi_{{}_{\textrm{int}}}^{(1)}(g=1;\phi) &
\\
=
\sum\limits_{\nu',s}
\int d^3 \boldsymbol{p}' d^3\boldsymbol{p} &
{\textstyle\frac{(m+\gamma^\mu p_\mu + \gamma^\mu p'_\mu)\gamma^{\nu'}u_s(\boldsymbol{p})\widetilde{\phi}(-|\boldsymbol{p}'|-p_0(\boldsymbol{p}), -\boldsymbol{p}'- \boldsymbol{p})}{(\boldsymbol{p}'|(\langle \boldsymbol{p}' | \boldsymbol{p} \rangle -|\boldsymbol{p}'|p_0(\boldsymbol{p}) )}}
& a_{\nu'}(\boldsymbol{p}') \, d_s(\boldsymbol{p})
\\
 + \sum\limits_{\nu',s}
\int d^3 \boldsymbol{p}' d^3\boldsymbol{p} & \ldots \,\,\,\,\,\,\,\,\,\,\,\,\,\,\,\,\,\,\,\,\,\,\,\,\,\,\,\,\,\,\,\,\,\,\,\,   
& a_{\nu'}(\boldsymbol{p}')^+ \, d_s(\boldsymbol{p})
\\
+ \sum\limits_{\nu',s}
\int d^3 \boldsymbol{p}' d^3\boldsymbol{p} & \ldots \,\,\,\,\,\,\,\,\,\,\,\,\,\,\,\,\,\,\,\,\,\,\,\,\,\,\,\,\,\,\,\,\,\,\,\, 
& a_{\nu'}(\boldsymbol{p}') \, d_s(\boldsymbol{p})^+
\\
+ \sum\limits_{\nu',s}
\int d^3 \boldsymbol{p}' d^3\boldsymbol{p} & \ldots \,\,\,\,\,\,\,\,\,\,\,\,\,\,\,\,\,\,\,\,\,\,\,\,\,\,\,\,\,\,\,\,\,\,\,\, 
& a_{\nu'}(\boldsymbol{p}')^+ \, d_s(\boldsymbol{p})^+
\end{align*}
where dots denote the kernels
\[
\kappa_{lm}(\phi)(\nu',\boldsymbol{p}', s,\boldsymbol{p})
= \pm {\textstyle\frac{(m+ \pm \gamma^\mu p'_\mu \pm )\gamma^{\nu'}u_s(\boldsymbol{p})\widetilde{\phi}(\pm |\boldsymbol{p}'|\pm p_0(\boldsymbol{p}), \pm \boldsymbol{p}'\pm \boldsymbol{p})}{(\boldsymbol{p}'|(\langle \boldsymbol{p}' | \boldsymbol{p} \rangle -|\boldsymbol{p}'|p_0(\boldsymbol{p}) )}},
\,\,\, l+m = 2
\]
with the respective $\pm$ sings in front of the whole expression and in front of the components $p'_{0}(\boldsymbol{p}') = |\boldsymbol{p}'|$,
$\boldsymbol{p}'$, $p_{0}(\boldsymbol{p}) = \sqrt{|\boldsymbol{p}|^2 + m^2}$, $\boldsymbol{p}$ of the momenta in the denominator, 
correspondingly to the annihilation or the creation operators. Using the elementary estimation
\[
\left | \textstyle{\frac{1}{(\boldsymbol{p}'|(\langle \boldsymbol{p}' | \boldsymbol{p} \rangle -|\boldsymbol{p}'|p_0(\boldsymbol{p}) )}} \right |
> \textstyle{\frac{1}{|\boldsymbol{p}'|^2(|p_0(\boldsymbol{p}) + |\boldsymbol{p}|)}}
\]
we see that the kernels $\kappa_{l,m}(\phi)(\nu',\boldsymbol{p}', s,\boldsymbol{p})$, regarded as two-particle functions of spin-momenta variables $(\nu',\boldsymbol{p}', s,\boldsymbol{p})$ do not belong to the tensor product of single particle Schwartz spaces or even to  the two-particle Hilbert
spaces having their $L^2(\mathbb{R}^{3\times 4} \times \mathbb{R}^{3\times 2})$-norms IR divergent. In particular $\psi_{{}_{\textrm{int}}}^{(1)}(g=1;\phi)$
acting on a finite number particle state with smooth Schwartz functions in each spin-momentum variable, lying in the domain $\mathcal{D}_0$, 
gives a nonnormalizable state which, moreover, is not smooth in  $(\boldsymbol{p}', \boldsymbol{p})$. Thus,  $\psi_{{}_{\textrm{int}}}^{(1)}(g=1)$
 cannot represent any operator valued distribution in the Wightman sense. But, as is easily seen, the above $\kappa_{lm}$, are kernels of well-defined
$\oplus_{1}^{4} \mathcal{E}^*$-valued distributions which, when integrated with 
$\xi_2 \otimes \xi_1 \in E_2 \otimes E^{\pm}_{1} = \mathcal{S}(\mathbb{R}^{3\times 4} \times \mathbb{R}^{3\times 2})$ 
regarded as functions of $(\nu, \boldsymbol{p}', s, \boldsymbol{p})$,
are continuous maps of $\xi_2 \otimes \xi_1$. Thus, by theorem \ref{Hida-ObataTheorem},  
$\psi_{{}_{\textrm{int}}}^{(1)}(g=1)$ is a finite sum of well-defined integral kernel operators in the sense of \cite{obataJFA}. 

In general 
higher order contributions to interacting fields in the adiabatic limit $g \rightarrow 1$ are not well-defined operator 
valued distributions in the Wightman sense and do not belong to class B) but only to class A), and this is the case only if the normalization in the splitting 
of causal distributions in the computation of the scattering operator is ``natural''. The adiabatic limit $g \rightarrow 1$
does not exist in the theory (I)-(V) based on Wightman distributions, so that in particular theorems \ref{g->1IntFields} and \ref{g->1IntFieldsm=0}
 of Subsection \ref{IntFields} cannot be proved within the approach based on Wightman distributions, contrary to what we have in the approach (I)-(V) based on the 
integral kernel operators in the sense of \cite{obataJFA}.

\subsection{UV and IR asymptotics}

For the proof of existence of IR and UV asymptotics of interacting fields, the application  of the Hida operators and integral kernel operators
is essential. Using the Hida operators in causal perturbative QED we can also compute the UV and IR asymptotics,
using the following facts (i)-(iv):
\begin{enumerate}

    \item[(i)] 
Each higher order contribution $A_{{}_{\textrm{int}}}^{(n)}$  to interacting e.m. potential field $A_{{}_{\textrm{int}}}$
exists as a generalized integral kernel operator $\Xi(\kappa_{lm})$
in the adiabatic limit $g\rightarrow 1$ in Bogoliubov's causal perturbative QED with Hida operators.

    \item[(ii)] 
The UV and  IR asymptotics should be $SL(2,\mathbb{C})$ invariant.

    \item[(iii)] The direct integral decomposition $\int U_\chi \, d\chi$ of the representation $U$ of 
$SL(2,\mathbb{C}) \subset T_4 \ltimes SL(2,\mathbb{C})$ acing in the full single particle Hilbert space
$\mathcal{H} = \int \mathcal{H}_\chi \, d\chi$ determines naturally direct integral 
decomposition $\int \Xi(\kappa_{\chi \, lm}) \, d\chi$ of $\Xi(\kappa_{lm}) = A_{{}_{\textrm{int}}}^{(n)}$.

    \item[(iv)] 
Decomposition components 
\begin{equation}\label{UVdecomposition}
A_{{}_{\chi \, \textrm{int}}}(x) = \sum \kappa_{\chi \, lm}(\ell_1,k_1, \ldots, \ell_{l+m}, k_{l+m}; x) a_{{}_{\chi \, \ell_1,k_1}}^{+}
\ldots a_{{}_{\chi \, \ell_{l+m}, k_{l+m}}}
\end{equation}
of $A_{{}_{\textrm{int}}}$ act in the Fock spaces $\Gamma(\mathcal{H}_{\chi})$ over the UV-asymptotcally  homogeneous states of UV-asymptotic
homogeneity degree determined by the decomposition parameter $\chi$.
\end{enumerate}

The generalized (discrete) integral kernel operators (\ref{UVdecomposition}) define the UV asymptotic parts of 
$A_{{}_{\textrm{int}}}$ of UV-asymptotic homogeneity degree determined by the decomposition parameter $\chi$
In order to find the IR asymptotics of $A_{{}_{\textrm{int}}}$ we need to compute the IR quasi-asymptotics of the scalar distributions
$\kappa_{\chi \, lm}(\ell_1,k_1, \ldots, \ell_{l+m}, k_{l+m}; x)$ in (\ref{UVdecomposition}) with respect to
the semigroup of scaling transformations $S_\lambda(x) = \lambda x$, $\lambda >0$. This is in general non-trivial and
uses the full theory of quasiasymptotics of distributions as given in \cite{Vindas,Vladimirov1,Vladimirov2}.
Collecting all $\kappa_{\chi \, lm}(\ell_1,k_1, \ldots, \ell_{l+m}, k_{l+m}; x)$ with common IR quasiasymptotic degree
we obtain the asymptotic part of $A_{{}_{\textrm{int}}}$  with fixed IR asymptotic degree. 

Let us give few words of explanation for (i)-(iv). The result (i) itself have been already briefly discussed in previous subsections. 
The kernels $\kappa_{\chi \, lm}(\phi)$ of (\ref{UVdecomposition}) evaluated at the space-time test function $\phi \in \oplus_{1}^{4} \mathcal{E}$,
are equal to the Fourier transforms $\mathcal{F}\big[\kappa_{lm}(\phi)\big](\chi, \ldots, \chi)$ of $\kappa_{\chi \, lm}(\phi)$, 
restricted to the diagonal, with the Fourier transform $\mathcal{F}$ associated to the
decomposition of the action $U^{\otimes \, (l+m)}$ of $SL(2,\mathbb{C})$ in the full $(l+m)$-particle test space $E^{\otimes \, (l+m)}$.
The existence of the adiabatic limit is essential, because in case we had the interaction $g\mathcal{L}$ with the modified intensity $g$
switched on, then the decomposition would depend non-trivially also on $g$.  
 
The problem of classification of all irreducible unitary representations of  $SL(2,\mathbb{C})$
has been completely solved by Gelfand and Naimark\cite{GelfandV, Geland-Minlos-Shapiro,NeumarkLorentzBook, nai1, nai2,nai3}, 
together with the problem of decomposition of tensor product of 
any unitary representations of  $SL(2,\mathbb{C})$ into irreducible components. But also the representation
of $SL(2,\mathbb{C})$ acting in the single particle space of the free e.m. potential field (in the Gupta-Bleuler gauge),
although not unitary but Krein-isometric, is decomposable with the decomposition which is determined by the 
normal scaling operator $S_\lambda$ and the respective decomposition components act in the Krein-Hilbert spaces
$\mathcal{H}_{\chi}$ of homogeneous states of homogeneity degree equal $\chi = -1 + i\nu$, $\nu \in \mathbb{R}$. Therefore
the kernels $\kappa_{\chi \, lm}$ in (\ref{UVdecomposition}) can indeed be computed explicitly, although the computation is
quite involved.

In order to explain the principle on which decomposition of $\Xi(\kappa_{lm})$, and of $\kappa_{lm}(\phi)$
associated with the decomposition of $SL(2,\mathbb{C})$ acting in $E^{\otimes \, (l+m)}$, is based, let us illustrate it in the simpler case
in  which we consider the standard unitary action of the translation group (analogue of $SL(2,\mathbb{C})$ group) on the Gelfand triple
\[
\mathcal{S}(\mathbb{R}; \mathbb{C}) \subset L^2(\mathbb{R}; \mathbb{C}) \subset \mathcal{S}(\mathbb{R}; \mathbb{C})^*
\]
 -- the analogue of the Gelfand triple:
\[
E^{\otimes \, (l+m)} \subset \mathcal{H}^{\otimes \, (l+m)} \subset E^{* \, \otimes \, (l+m)}.
\]
Decomposition of the unitary action $U$ of the translation group into irreducible components $U_\chi$ determines 
the decomposition $L^2(\mathbb{R}) = \int \mathcal{H}_{{}_{\chi}} \, d\chi$ into one dimensional Hilbert spaces $\mathcal{H}_{{}_{\chi}} = \mathbb{C}$,
and with the spectrum of decomposition equal to $\mathbb{R}$ with the spectral measure $d\chi$ of decomposition equal
to the ordinary Lebesgue measure. Therefore each $f \in L^2(\mathbb{R})$, and in particular each $f\in \mathcal{S}(\mathbb{R})$
has unique decomposition with decomposition components 
\[
\big(f\big)_{{}_{\chi}} = \mathcal{F}f(\chi), \,\,\, \chi \in \mathbb{R},
\]
where $\mathcal{F}$ is the ordinary Fourier transform, associated to the decomposition of the unitary action of the translation
group on $\mathbb{R}$. Decomposition of the elements $f$ of the nuclear space $\mathcal{S}(\mathbb{R})$ determines decomposition
of a (decomposable) distribution $F \in \mathcal{S}(\mathbb{R}; \mathbb{C})^*$
by the following canonical formula 
\begin{multline*}
\langle F, f\rangle = \int\limits_{\textrm{Spec} = \mathbb{R}} F_{{}_{\chi}} \mathcal{F}{f}(\chi) \, d \chi =
\int\limits_{{}_{\textrm{Spec} = \mathbb{R}}} \big\langle F_{{}_{\chi}}, \mathcal{F}{f}(\chi)\big\rangle \, d \chi
\\
=
\int\limits_{{}_{\textrm{Spec} = \mathbb{R}}} \big\langle \mathcal{F}F(\chi), (f)_{{}_{\chi}}\big\rangle \, d \chi
, \,\,\,\,\,\,
f \in \mathcal{S}(\mathbb{R}; \mathbb{C}).
\end{multline*}
We thus see that decomposable $F$ whose decomposition measures $d\chi$ are absolutely continuous with respect to the Lebesgue
measure, are precisely those distributions whose Fourier transforms are regular function-like distributions. But more generally,
$F$ is decomposable, with any fixed $\sigma$-measure $d\chi$ on $\mathbb{R}$ iff its Fourier transform is equal to a 
$\sigma$-measure on $\mathbb{R}$.

The same situation we have for decomposable $\kappa_{lm}(\phi)$ which can similarly be decomposed 
with decomposition associated with the decomposition of $SL(2,\mathbb{C})$ acting in $E^{\otimes \, (l+m)}$
and which determine decomposition of the associated integral kernel operator. Namely, we put 

\begin{definition}
Let
\[
\Phi,\Psi \in (E),
\,\,\,\, \Phi = \int \Phi_{{}_{\chi}} \, d \chi,
\,\,\,\, \Psi = \int \Psi_{{}_{\chi}} \, d \chi
\]
be any two elements of the test Hida space with their direct integral decompositions. Let $d \chi$
be a $\sigma$-measure on the spectrum of the decomposition of the representation of $SL(2, \mathbb{C})$ acting
in the single particle Hilbert space $\mathcal{H}$. We say that the generalized
integral kernel operator $\Xi(\kappa_{lm})$ is equal to the direct integral
\[
\Xi(\kappa_{lm}) = \int \Xi_{{}_{\chi}}(\kappa_{\chi \, lm}) \, d \chi
\]
of (discrete-) integral kernel operators $\Xi_{{}_{\chi}}(\kappa_{\chi \, lm})$,
acting in the Fock spaces over the single particle Gelfand triples
$E_{{}_{\chi}} \subset \mathcal{H}_{{}_{\chi}} \subset E_{{}_{\chi}}^{*}$, if
\begin{multline*}
\int \big\langle\big\langle \Xi_{{}_{\chi}}\big(\kappa_{\chi \, lm}(\phi)\big)
\Phi_{{}_{\chi}}, \Psi_{{}_{\chi}} \big\rangle\big\rangle \, d \chi
= \int \Big\langle \kappa_{\chi \, lm}(\phi), \big(\eta_{{}_{\Phi,\Psi}}\big)_{{}_{\chi}} \Big\rangle \, d \chi
\\
=\langle \kappa_{lm}(\phi), \eta_{{}_{\Phi,\Psi}} \rangle
=\big\langle \big\langle \Xi(\kappa_{lm}(\phi))\Phi, \Psi \big\rangle \big\rangle
\end{multline*}
for all
\[
\Phi,\Psi \in (E), \phi \in \mathcal{E}.
\]
\qed
\label{decopositionOfXi}
\end{definition}

Integral kernel operator $\Xi(\kappa_{lm}(\phi))$, being uniquely determined by its kernel $\kappa_{lm}(\phi)$,
has the decomposition canonically associated to the decomposition of the distributional kernel
$\kappa_{lm}(\phi)$. Similar decomposition does not make any sense for the operator valued distributions
in the sense of Wightman.  

A quantum theory of classical homogeneous of degree $-1$ e.m. potential field has been developed long time ago
by Staruszkiewicz\cite{Staruszkiewicz,Staruszkiewicz1992ERRATUM, Staruszkiewicz2009}. The IR asymptotics of $A_{{}_{\textrm{int}}}$
of degree $\chi=-1$, computed as above, exists and is non-trivial (nonzero) for QED. It coincides with
the theory of Staruszkiewicz \cite{Staruszkiewicz,Staruszkiewicz1992ERRATUM, Staruszkiewicz2009} if and only if 
we add the following assumption: for the total charge operator $Q$, acting in the (asymptotically) homogeneous states
of the free fields coupled to the electromagnetic potential, there exist the phase operator $S_0$ which provides,
together with $Q$, the spectral realization of the gauge group $U(1)$ in the sense of Connes \cite{Connes_spectral}. 
This additional assumption
emerges from the comparison of the IR (asymptotically) homogeneous part of degree $-1$ with the theory
of  Staruszkiewicz \cite{Staruszkiewicz,Staruszkiewicz1992ERRATUM, Staruszkiewicz2009} and accounts for the
universality of the electric charge, which cannot be explained solely within the Bogoliubov's causal
QED with Hida operators, as in principle we can put different coupling constants for different charged fields.

\section*{Acknowledgments}

The author would like to express his deep gratitude to Professor I. Volovich and Professor D. Kazakov 
for the very helpful discussions. I am grateful to Professor Volovich who motivated me
for giving rigorous analysis of the Bogoliubov's causality axioms with Hida operators. 
I am grateful to Prof. D. Kazakov for bringing the problem, solution of which
is summarized in theorem \ref{g->1IntFieldsm=0}, to my attention.
He also would like to thank for the excellent conditions for work at JINR, Dubna.
He would like to thank Professor M. Je\.zabek 
for the excellent conditions for work at INP PAS in Krak\'ow, Poland
and would like to thank Professor A. Staruszkiewicz and 
Professor M. Je\.zabek for the warm encouragement.


\begin{thebibliography}{10}

\bibitem{Bogoliubov_Shirkov} 
N. N. Bogoliubov, N. N., D. V. Shirkov, D. V., {\em Introduction to the Theory of Quantized Fields}, 2nd edn. 
(John Wiley \& Sons, Inc., New York, Chichester, Brisbane, Toronto, 1980; first Eng. edn.: 1959, first Russian edn.: 1956).

\bibitem{Epstein-Glaser} 
H. Epstein, V. Glaser, The role of locality in perturbation theory, {\em Ann. Inst. H. Poincar\'e} {\bf A19}, 211 (1973).

\bibitem{wig} 
R. F. Streater and A. S. Wightman, {\em PCT, Spin and Statistics, and All
That}, (W. A. Benjamin, Inc., New York, 1964).  

\bibitem{obataJFA} 
N. Obata, Operator calculus on vector-valued white noise functionals, {\em J. of Funct. Anal.} {\bf 121}, 185-232 (1994).

\bibitem{GelfandIV} 
I. M. Gelfand and N. Ya. Vilenkin, {\em Applications of Harmonic Analysis: Generalized functions. Vol. 4} 
(Acad. Press, New York, 1964).

\bibitem{obata} 
N. Obata, 
An analytic characterization of symbols of
operators on white noise functionals,
{\em J. Math. Soc. Japan} {\bf 45}, 421 (1993).

\bibitem{Schaefer} 
H. H. Schaefer, {\em Topological vector spaces}, 2nd ed.  (Springer, New York 1999). 

\bibitem{Scharf} 
G. Scharf, {\em Finite Quantum electrodynamics} (Dover Publications, Mineola, New York, 2014).

\bibitem{Berezin}  
F. A. Berezin, {\em The method of second quantization} (Acad. Press, New York, London, 1966).


\bibitem{Vindas} 
J. Vindas, Structural theorems for quasiasymptotics of distributions at infinity, {\em Publications De'Institut Math\'ematique}
{\bf 84}, 159 (2008).

\bibitem{Vladimirov1}
V. S. Vladimirov, Yu. N. Drozhzhinov, B. I. Zav’yalov, Tauberian theorems for generalized
functions in a scale of regularly varying functions and functionals, dedicated to Jovan Kara-
mata, {\em Publ. Inst. Math. (Beograd)} {\bf 71}, 123 (2002) (in Russian).

\bibitem{Vladimirov2}
V. S. Vladimirov, Yu. N. Drozzinov, B. I. Zavialov, {\em Tauberian Theorems for Generalized Func-
tions} (Kluwer, Maine, 1988).

\bibitem{Geland-Minlos-Shapiro} 
I. M. Gelfand, R. A. Minlos, Z. Ya. Shapiro, {\em Representations of the rotation and Lorentz groups and their applications} 
(Pergamon Press Book, The Macmillan Company, New York, 1963).

\bibitem{NeumarkLorentzBook} 
M. A. Naimark, {\em Linear representations of the Lorentz group} 
(Pergamon Press, Oxford, London, Edinburgh, New York, Paris, Frankfurt, 1964).

\bibitem{nai1} 
M. A. Naimark, 
Decomposition of a tensor product of
irreducible representations of the proper Lorentz group into
irreducible representations. I. The case of a tensor product
of representations of the fundamental series, 
{\em Tr. Mosk. Mat.
Obs.} {\bf 8} 121 (1959). 

\bibitem{nai2}
M. A. Naimark, 
Decomposition of a tensor product of
irreducible representations of the proper Lorentz group into
irreducible representations. II. The case of a tensor product of
representations of the fundamental and complementary series,
{\em Tr. Mosk. Mat.
Obs.} {\bf 9} 237 (1960). 

\bibitem{nai3} 
M. A. Naimark, 
Decomposition of a tensor product of
irreducible representations of the proper Lorentz group into
irreducible representations. III. The case of a tensor product
of representations of the supplementary series, 
{\em Tr. Mosk. Mat.
Obs.} {\bf 10}, 181 (1961).

\bibitem{GelfandV} 
I. M. Gelfand, M. I. Graev, and N. Ya. Vilenkin, {\em Generalized Functions. Vol V}
(Academic Press, New York and London, 1966).

\bibitem{Staruszkiewicz} 
A. Staruszkiewicz, Quantum mechanics of phase and charge and quantization of the Coulomb field,
{\em Ann. Phys. (N.Y.)} {\bf 190}, 354 (1989).

\bibitem{Staruszkiewicz1992ERRATUM} 
A. Staruszkiewicz, The quantized Coulomb field and irreducible unitary representations 
of the proper, ortochronous Lorentz group,
{\em Acta Phys. Polon.} {\bf B23}, 591 (1992).

\bibitem{Staruszkiewicz2009} 
A. Staruszkiewicz, A new proof of existence of a bound state in the quantum Coulomb field II, 
{\em Reports on Math. Phys.} {\bf 64}, 293 (2009).

\bibitem{Connes_spectral} 
A. Connes, On the spectral characterization of manifolds, {\em J. Noncommutat. Geom.} {\bf 7}, 1 (2013).






\end{thebibliography}
\end{document}